\documentclass[letterpaper, paper,10pt]{AAS}	

\usepackage{bm}
\usepackage{amsmath}
\usepackage{amssymb}
\usepackage{gensymb}
\usepackage{relsize}
\usepackage{placeins}
\usepackage{adjustbox}
\usepackage{graphicx}
\usepackage{caption}
\usepackage{subcaption}

\usepackage[colorlinks=true, pdfstartview=FitV, linkcolor=black, citecolor= black, urlcolor= black]{hyperref}
\usepackage{algorithm}
\usepackage{algpseudocode}

\DeclareMathOperator*{\argmin}{arg\,min}

\PaperNumber{24-456}

\begin{document}


\title{End-to-End Lyapunov-Based Eclipse-Feasible Low-Thrust Transfer Trajectories to NRHO}

\author{Nicholas P. Nurre\thanks{Graduate Student, Department of Aerospace Engineering, Auburn University, 141 Engineering Dr Auburn, AL 36849.}\, and  
Ehsan Taheri\thanks{Assistant Professor, Department of Aerospace Engineering, Auburn University, 141 Engineering Dr Auburn, AL 36849.}}

\maketitle{} 		

\begin{abstract}
Generating low-thrust transfer trajectories between Earth and the Near Rectilinear Halo Orbit (NRHO), that is selected for NASA's Gateway, can be challenging due to the low control authority available from the propulsion system and the important operational constraint that the duration of all eclipses has to be less than a prescribed 90-minute threshold. We present a method for generating eclipse-feasible, minimum-time solutions to the aforementioned trajectory design problem using a Lyapunov control law. Coasting is enforced during solar eclipses due to both the Earth and Moon. We used particle swarm optimization to optimize the NRHO insertion date, time of flight, and control law parameters according to a cost function that prioritizes 1) convergence to the target orbit, 2) satisfaction of eclipse-duration constraints, and 3) minimization of time of flight. Trajectories can serve as initial guesses for NASA's high-fidelity trajectory design tools such as Copernicus and GMAT.
\end{abstract}

\section{Introduction}
Design of low-thrust transfers to the vicinity of the Moon is of interest, with the selection of a Near-Rectilinear Halo Orbit (HALO) by NASA as a staging platform for exploration of the Moon and beyond \cite{david_e_lee_white_2019}. Designing low-thrust trajectories can be quite difficult due to the combination of the very low control authority available from low-thrust propulsion systems and the highly nonlinear dynamical environment of the cislunar region. Since existing low-thrust spacecraft are powered with solar arrays, it's essential that no solar eclipses last longer than a certain time interval. For instance, for the transfer of the Co-Manifested Vehicle (CMV) of Gateway, all eclipse durations need to be less than 90 minutes \cite{mcguire_overview_2021}. Extended operation of spacecraft within eclipses can deplete the batteries and lead to a complete loss of the vehicle.

Due to low propulsive accelerations on the order of $1.0\times10^{-4}$ m/s$^2$, transfer trajectories can require times of flight on the order of a year or longer. Furthermore, low-thrust trajectories consist of different phases, where the primary gravitational influence shift from Earth to the Moon.  Therefore, transfer trajectories are typically solved in multiple subphases. Ref. \cite{mcguire_overview_2021} gives an overview of NASA's third and most recent Design Reference Mission (DRM) for the transfer of the CMV, which was designed in four subphases using Copernicus \cite{ocampo_architecture_2003}. Indirect optimization methods are also used for designing low-thrust trajectories to quasi-periodic, near-rectilinear Halo orbits that leverages ephemeris-driven, “invariant manifold analogs” as long-duration asymptotic terminal coast arcs \cite{singh2021eclipse}. All discontinuous events (such as entry into and exit from Earth eclipses and throttle switches) are made smooth through the powerful and novel Composite Smooth Control (CSC) framework \cite{taheri2020novel}. Ref. \cite{patrick_hybrid_2023} solves a similar transfer problem in two subphases with an indirect method to optimize a powered Earth-spiral subphase that is then heuristically patched into a second ballistic subphase. Numerical continuation and homotopy methods are fundamental to the convergence of the Hamiltonian two-point boundary-value problems associated with indirect methods \cite{taheri2018generic,ayyanathan2022mapped,kovryzhenko2023vectorized,taheri2023l2,saloglu2024acceleration}. 
To consolidate the design approach, we solved a similar problem in one phase (i.e., an Earth-centered perturbed two-body model is used with perturbations due to the Moon, Sun, and Earth's second zonal harmonic subject to Earth eclipses) with a multiple-shooting indirect method  \cite{nurre_end--end_2024}. Both minimum-time and minimum-fuel solutions were achieved by starting with a high level of spacecraft acceleration and performing numerical continuation to gradually reduce the value of acceleration. 

Another approach for designing low-thrust many-revolution trajectories is to use Lyapunov-based approaches. Lyapunov control (LC) is based on the Lyapunov stability theory, using which an LC law is obtained by finding the expression for control which makes the time derivative of a control-Lyapunov function (CLF) for the system negative \cite{noble_ariel_hatten_critical_2012,schaub_analytical_2018,epenoy_lyapunov-based_2019}. 
The CLF is positive in terms of the states of the system and should become 0 at the equilibrium (corresponding to a desired ``goal'' or target state). The method can be likened to converting the second-order trajectory optimization problem into a first-order stabilization problem. We note that LC has been used extensively for low-thrust trajectory optimization \cite{\cite{petropoulos_simple_2003}}. For instance, Ref. \cite{shannon_rapid_2020} applies Q-law \cite{petropoulos_simple_2003} to solve Earth-Moon transfers in two subphases. The results are shown to serve as high-quality initial guesses for GPOPS-II \cite{patterson_gpops-ii_2014} -- a pseudospectral direct method solver. The authors leverage the computational efficiency of LC to perform an extensive trade study over potential epochs and departure orbits, allowing for an \textit{a posteriori} analysis of the eclipses. Ref. \cite{epenoy_lyapunov-based_2019} proposes a hybrid LC based methodology for designing Earth-Moon transfers in a full-ephemeris model. A study on the sensitivity of the LC law to missed-thrust events is also performed to demonstrate the robustness of the control law. 

In this paper, we consider a single-phase design approach similar to what we considered in Ref. \cite{nurre_end--end_2024}; however, a Lyapunov control (LC) law is used instead of an indirect method to solve the trajectory optimization problem. 
Further, convergence of LC laws is asymptotic and depends on the rate at which the Lyapunov function value is decreased. Finite-time convergence to the goal can be achieved by parameterizing the CLF and optimizing these new parameters with respect to a cost function to obtain near-optimal solutions (for example, with respect to flight time or fuel consumption). In addition, convergence tolerances can be set that define when the propagated state is ``close enough'' to the goal/target state. LC laws are also straightforward to design and implement, and more importantly, are closed-loop in nature (i.e., they only depend on the current state) \cite{ozimek2023onboard}. Thus, a motivation for this work is to rapidly solve low-thrust transfer problems. These solutions, in turn, can be used as initial guesses for other high-fidelity trajectory optimization tools that will provide more optimal solutions that precisely satisfy boundary conditions like, for instance, the indirect method in Ref. \cite{nurre_end--end_2024} or Copernicus. 

An important operational constraint, for low-thrust trajectories to the Gateway, is that the duration of all eclipses has to be less than a prescribed 90-minute threshold. Mission design strategies for ensuring all solar eclipse durations are less than the prescribed time often entail generating a large set of reference trajectories for a range of departure epochs, as was done for Artemis I \cite{williams_new_2023}, to have a variety of options. However, many departure windows may not be feasible. Ref. \cite{williams_new_2023} reports that 18\% of all launch days were infeasible for Artemis I. The eclipse-duration constraint can pose more challenges for extremely low-thrust propulsion systems that require significantly longer times of flight. For low-thrust transfers departing from a Geostationary Transfer Orbit (GTO) to the Moon, Earth eclipse events highly depend on the departure epoch and GTO orientation. Adjustments can be made to these values to mitigate eclipse durations. However, even with an analyst's extensive experience, this post-processing approach can take many iterations to identify feasible launch opportunities. Long intermittent Earth and Moon eclipses occurring in cislunar space, when the spacecraft's relative velocity is much slower, are possible and not as easily preventable.

Incorporating eclipse-duration constraints \textit{within} the trajectory optimization can could increase the number of feasible departure windows and improve optimality by allowing the trajectory optimization to be ``aware'' of such constraints within the optimization process. Eclipse-duration constraints are difficult to enforce due to the fact that 1) the number of eclipses is not known \textit{a priori} and 2) the number and duration of eclipses can change within the trajectory optimization process. Ref. \cite{williams_new_2023} outlines an effective strategy that treats eclipse durations as inequality constraints in Copernicus. The results, in the paper, indicate that it was possible to increase the number of feasible launch dates by about 20\% for Artemis I. In this paper, however, we attempt to satisfy the eclipse maximum-duration constraint with a soft-penalization enforced while optimizing the parameters of the transfer problem. This eliminates an analyst-in-the-loop design approach and automates and facilitates the solution procedure.

The contributions of the paper are as follows. A LC law is derived and used to solve transfer trajectory optimization problems similar to those in Refs. \cite{mcguire_overview_2021,patrick_hybrid_2023}. The LC law can only be used to transfer the spacecraft starting from a fully defined state into an orbit, i.e., in its standard formulation LC cannot be used for rendezvous type transfers unless modifications are applied to the problem formulation \cite{narayanaswamy2023equinoctial}. Therefore, always starting from a point on the NRHO, the control law is applied backward in time for a departure from a GTO to an insertion at NRHO and forward in time for a departure from NRHO to an insertion at GTO. We only consider transfer maneuvers from GTO to NRHO. The GTO departure time and true anomaly are free and the NRHO insertion time is fixed. However, we consider the NRHO insertion time to be a design parameter. Because the ephemeris-propagated and ephemeris-corrected NRHO provided in Ref. \cite{david_e_lee_white_2019} is considered, the entire state on the NRHO can be defined by the time (i.e., epoch). While numerically integrating the spacecraft equations of motion, event detection is used to determine if the solutions converge, if the spacecraft intersects the surface of the Earth or Moon, and when eclipses due to the Earth and Moon occur. Coasting is enforced during eclipses and the duration of each eclipse is calculated. Particle swarm optimization \cite{kennedy1995particle} is used to optimize the NRHO insertion date, time of flight, and parameters of the CLF with respect to a hierarchical cost function. The cost function prioritizes 1) convergence to the target orbit, 2) satisfaction of the maximum-eclipse-duration constraint, and 3) minimization of the time of flight.


\section{Dynamical Model}
The entire transfer problem is solved in the J2000 Earth-centered inertial (ECI) frame.
All accelerations are expressed with respect to this frame. The spacecraft's motion is modeled with position, \(\bm{r}=[x,\,y,\,z]^\top\), and velocity, \(\bm{v}=[v_x,\,v_y,\,v_z]^\top\), vectors. The spacecraft's state vector is \(\bm{x}=[\bm{r}^\top,\,\bm{v}^\top]^\top\) and the equations of motion are defined as,
\begin{equation} \label{eq: eom}
    \dot{\bm{x}} = \bm{f}\left(t,\bm{x},\hat{\bm{\alpha}}\right) = \begin{bmatrix} \bm{v} \\ \bm{a}_\text{kep} + \bm{a}_\text{3rd} + \bm{a}_{J_2} + \hat{\bm{\alpha}} a_\text{sc} \delta_\text{ecl} \end{bmatrix},
\end{equation}
where \(t\) denotes time, \(\bm{a}_\text{kep}\) is the two-body (Keplerian) acceleration due to the Earth, \(\bm{a}_\text{3rd}\) is the collection of third-body gravitational perturbations, and \(\bm{a}_{J_2}\) is the acceleration due to Earth's \(J_2\) gravitational perturbation. In the last acceleration term, $\hat{\bm{\alpha}} a_\text{sc} \delta_\text{ecl}$, which denotes the acceleration produced by the propulsion system, \(\hat{\bm{\alpha}}\) is the thrust steering unit vector and \(\delta_\text{ecl}\in\{0,1\}\) is the eclipse-triggered throttle factor. Since the contribution and emphasis of the work is on satisfying the maximum-eclipse-duration constraint, a constant spacecraft acceleration is assumed with its value set to \(a_\text{sc}=1.0\times10^{-4}\) m/s\(^2\). This value is chosen to match the transfer problems in Refs. \cite{mcguire_overview_2021,patrick_hybrid_2023}. Propellant-mass considerations belong to our future work. The thrust steering unit vector can freely orient in space, but is constrained to a unit vector, i.e.,
\begin{equation} \label{eq: control constraint}
    \hat{\bm{\alpha}}^\top\hat{\bm{\alpha}} = 1.
\end{equation}

In this work, the change in spacecraft mass is not taken into account. But, our future work will investigate implementing a LC law coasting mechanism to obtain suboptimal minimum-fuel solutions, like the one that is introduced with Q-law in Ref. \cite{petropoulos_low-thrust_2004}. Earth's two-body acceleration can be written as,
\begin{equation} \label{eq: two body}
    \bm{a}_\text{kep} = -\frac{\mu_\text{Earth}}{r^3}\bm{r},
\end{equation}
where \(r=\|\bm{r}\|\) and \(\mu_\text{Earth}\) is the gravitational parameter of the Earth. Perturbing accelerations due to the gravity of the Moon, Sun, and Jupiter are considered and written as \cite{battin_introduction_1999,betts_optimal_2003},
\begin{equation} \label{eq: 3rd body accelerations new}
    \bm{a}_{\text{3rd}} = -\mathlarger{\mathlarger{\sum}}_{k\in K} \mu_k\frac{\bm{r} + F\left(q_k\right)\bm{r}_k}{\|\bm{r} - \bm{r_k}\|^3},
\end{equation}
where
\begin{align}
    F\left(q_k\right) & = q_k\left(\frac{3 + 3q_k + q_k^2}{1+\left(1+q_k\right)^{3/2}}\right), & q_k & = \frac{\bm{r}^\top\left(\bm{r}-2\bm{r}_k\right)}{\bm{r}_k^\top\bm{r}_k},
\end{align}
with \(K\in\left\{\text{Moon},\,\text{Sun},\,\text{Jupiter}\right\}\), and \(\bm{r}_k\) denotes the position of the \(k\)-th body with respect to the Earth. Note that this formulation avoids any numerical error due to cancellations when terms are of significantly different values \cite{battin_introduction_1999}. The acceleration vector due to Earth's \(J_2\) gravitational perturbation is written as \cite{schaub_analytical_2018,sowell2024eclipse},
\begin{align}
    \bm{a}_{J_2} = \frac{3J_2\mu_\text{Earth} R_\text{Earth}^2}{2r^4}\begin{bmatrix} \frac{x}{r}\left(5\frac{x^2}{r^2}-1\right), & \frac{y}{r}\left(5\frac{y^2}{r^2}-1\right), & \frac{z}{r}\left(5\frac{z^2}{r^2}-3\right) \end{bmatrix}^\top,
\end{align}
where \(R_\text{Earth}\) is the mean radius of the Earth and \(J_2=1082.63\times10^{-6}\). 

A canonical distance unit (DU) is defined by one Earth radius, \(R_\text{Earth}\), and a canonical time unit (TU) is defined such that the scaled value of Earth's gravitational parameter is 1 DU$^3$/TU$^2$. These canonical distance and time units are then used to scale all states and parameters of the dynamical model. Future work could include investigating more sophisticated scaling methods and even time regularization methods such as Ref. \cite{leith_time_2023} that might make numerical integration of Eq.~\eqref{eq: eom} more efficient. All planetary ephemerides and constants are obtained using NASA's SPICE toolkit \cite{acton_ancillary_1996} and the generic kernel files\footnote{\url{https://naif.jpl.nasa.gov/pub/naif/generic_kernels/}} \[\verb|de440.bsp|,~\verb|naif0012.tls|,~\verb|pck00011.tpc|,~\text{and}~\verb|gm_de440.tpc|.\] 

It is computationally inefficient to call SPICE routines during numerical integration. Thus, all ephemerides obtained positions, i.e., \(\bm{r}_\text{k}\) appearing in Eq.~\eqref{eq: 3rd body accelerations new} and in the eclipse model presented in the next section, are fitted by a spline function, which has proved to be more computationally efficient. The interpolation is performed to an accuracy on the order of 0.1 m.

\section{Eclipse Model}
\begin{figure}[h]
    \centering
    \includegraphics[width=\textwidth]{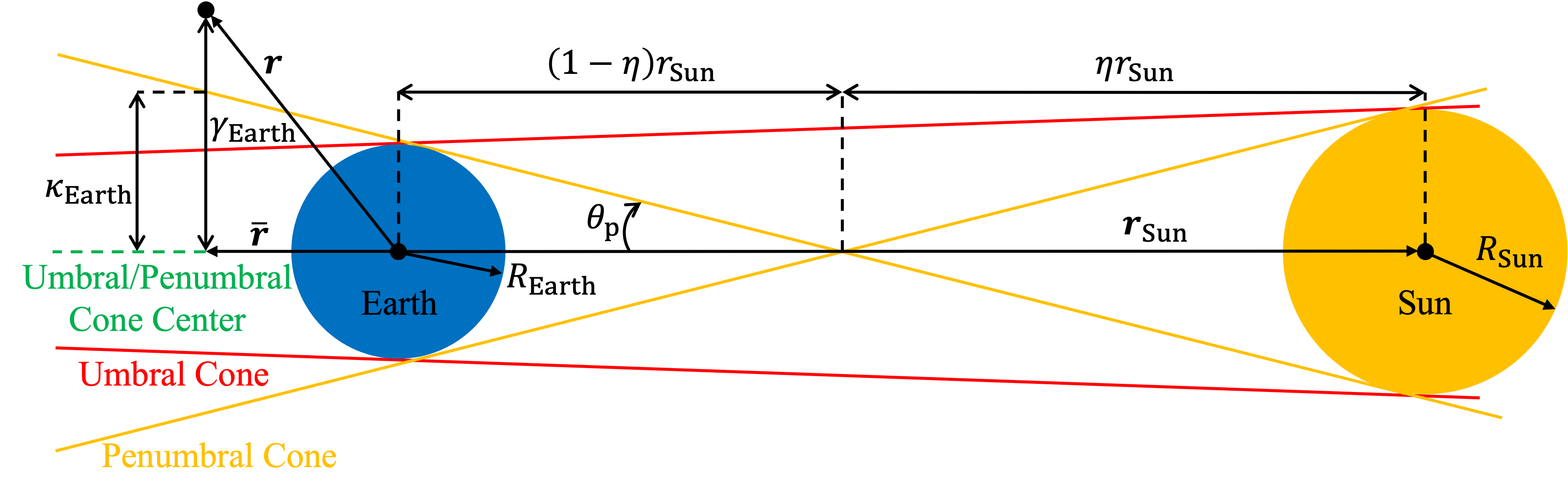}
    \caption{Sun-Earth eclipse geometry.}
    \label{fig: eclipse geometry}
\end{figure}

In this work, eclipses due to the Earth and Moon are considered. The cylindrical eclipse model from Ref. \cite{betts_optimal_2015} is used. The eclipse model assumes the Earth, Moon, and Sun to be perfect spheres and the spacecraft to be a point mass. Eclipse coasting is enforced during both umbra (total eclipse) and penumbra (partial eclipse). Thus, only the penumbra exits and entries are calculated, since umbra occurs inside penumbra. Figure \ref{fig: eclipse geometry} illustrates the Sun-Earth shadow geometry. Note that Figure \ref{fig: eclipse geometry} is greatly exaggerated and not drawn to scale. The same geometry is also used for modeling Moon eclipses.

Let \(r_\text{Sun}=\|\bm{r}_\text{Sun}\|\) be the distance between the Earth and the Sun and let \(\eta \in [0,1]\). From the geometrical proportion of the penumbral cone, we have
\begin{equation}
    \frac{\left(1-\eta\right)r_\text{Sun}}{2R_\text{Earth}}=\frac{\eta r_\text{Sun}}{2 R_\text{Sun}},
\end{equation}
where the value \(\eta\) can be expressed as,
\begin{equation}
    \eta = \frac{R_\text{Sun}}{R_\text{Earth}+R_\text{Sun}}.
\end{equation}

Therefore, the angle that the penumbral cone makes with respect to \(\bm{r}_\text{Sun}\) can be expressed as,
\begin{equation}
    \theta_\text{p} = \sin^{-1}{\left(\frac{R_\text{Earth}}{\left(1-\eta\right)r_\text{Sun}}\right)} = \sin^{-1}{\left(\frac{R_\text{Earth} + R_\text{Sun}}{r_\text{Sun}}\right)}.
\end{equation}

The position of the spacecraft projected onto \(\bm{r}_\text{Sun}\) is defined as,
\begin{equation}
    \bar{\bm{r}} = \left(\bm{r}^\top\hat{\bm{r}}_\text{Sun}\right)\hat{\bm{r}}_\text{Sun},
\end{equation}
where \(\hat{\bm{r}}_\text{Sun}=\bm{r}_\text{Sun}/r_\text{Sun}\). Earth shadows only occur on the side of the Earth opposite the Sun, i.e., the spacecraft can only encounter a shadow when \(\bm{r}^\top\hat{\bm{r}}_\text{Sun}<0\). Let the distance between the spacecraft and the center of the penumbral cone be defined as,
\begin{equation} \label{eq: gamma earth}
    \gamma_\text{Earth} = \left\|\bm{r} - \bar{\bm{r}}\right\|,
\end{equation}
and let the distance between the penumbral terminator and the center of the penumbral cone at the projected spacecraft location be defined as,
\begin{equation} \label{eq: kappa earth}
    \kappa_\text{Earth} = \left(\left(1-\eta\right)r_\text{Sun} + \|\bar{\bm{r}}\|\right)\tan{\left(\theta_\text{p}\right)}.
\end{equation}

Therefore, it can be stated that the spacecraft is in the Earth shadow when \(\gamma_\text{Earth} < \kappa_\text{Earth}\) and \(\bm{r}^\top\hat{\bm{r}}_\text{Sun}<0\) and not in a shadow otherwise. 

Let \(\gamma_\text{Moon}\) and \(\kappa_\text{Moon}\) denote the same definitions as Eqs.~\eqref{eq: gamma earth} and \eqref{eq: kappa earth}, respectively, but for the Moon-Sun eclipse model. Also, let \(\bm{r}_\text{sc/Moon}=\bm{r} - \bm{r}_\text{Moon}\) be the position of the spacecraft with respect to the Moon and \(\hat{\bm{r}}_\text{Sun/Moon} = \left(\bm{r}_\text{Sun} - \bm{r}_\text{Moon}\right)/\left\|\bm{r}_\text{Sun} - \bm{r}_\text{Moon}\right\|\) be the unit vector pointing towards the Sun with respect to the Moon. The following switching functions can then be defined
\begin{align}
    S_\text{ecl,Earth,1} & = \gamma_\text{Earth} - \kappa_\text{Earth} , & S_\text{ecl,Earth,2} & = \bm{r}^\top\hat{\bm{r}}_\text{Sun}, \label{eq: earth ecl switching functions} \\
    S_\text{ecl,Moon,1} & = \gamma_\text{Moon} - \kappa_\text{Moon} , & S_\text{ecl,Moon,2} & = \bm{r}_\text{sc/Moon}^\top\hat{\bm{r}}_\text{Sun/Moon}. \label{eq: moon ecl switching functions}
\end{align}

The eclipse-triggered throttle factor in Eq.~\eqref{eq: eom}, \(\delta_\text{ecl}\), can be defined as a multiplication of two factors as,
\begin{equation} \label{eq: ecl function}
    \delta_\text{ecl} = \delta_\text{ecl,Earth}\times\delta_\text{ecl,Moon},
\end{equation}
where
\begin{equation}
    \delta_\text{ecl,Earth} = \begin{cases} 0, &    S_\text{ecl,Earth,1}<0~\text{and}~S_\text{ecl,Earth,2}<0, \\ 1, & \text{else}, \end{cases}
\end{equation}
and 
\begin{equation}
    \delta_\text{ecl,Moon} = \begin{cases} 0, &    S_\text{ecl,Moon,1}<0~\text{and}~S_\text{ecl,Moon,2}<0, \\ 1, & \text{else}. \end{cases}
\end{equation}

During numerical integration of Eq.~\eqref{eq: eom}, an event-detection algorithm is used to determine the exact time of each eclipse entry, \(t_{\text{ecl},n}^-\), and exit, \(t_{\text{ecl},n}^+\), for all $n$ eclipses with \(n=1,\dots,N\). The duration of each eclipse, denoted as \(d_{\text{ecl},n} = t_{\text{ecl},n}^+ - t_{\text{ecl},n}^-\), is also calculated. Note that the total number of Earth and Moon eclipses, \(N\), is not known in advance. The eclipse constraint, that all eclipses be less than 90 minutes \cite{mcguire_overview_2021}, is considered in this work and can be expressed formally as,
\begin{equation} \label{eq: ecl constraint}
    d_{\text{ecl},n} \leq 90~\text{minutes},~\forall~n = 1,\dots,N.
\end{equation}

Unlike the eclipse model used by the authors in Ref.~\cite{nurre_end--end_2024}, the domain for the eclipse model from Ref. \cite{betts_optimal_2015} is defined interior to the occulting bodies. This was the main reason that we adopted this eclipse model. While optimizing all the parameters of the transfer problem, event detection is used to stop integration when the spacecraft intersects the Earth or Moon. This logic works most of the time, however, sometimes the dynamics become significantly nonlinear around the Moon and the event-detection algorithm misses the intersection event. When the model from Ref. \cite{aziz_low-thrust_2018} is used, \verb|NaN|'s are returned since the model is undefined for the domain inside the occulting body, which breaks the optimization routine. A similar problem is reported with the eclipse model used in \cite{williams_new_2023} along with a strategy for overcoming it. The previously presented eclipse model circumvents this problem altogether.

\section{Trajectory Optimization Problem}
The transfer problem, departing from a GTO at a date \(t_0\) and inserting into NRHO at a later date \(t_f\), is solved backwards in time over the time horizon 
\begin{align} \label{eq: time horizon}
    &t\in[t_f,t_0], & & t_f > t_0.
\end{align}

The GTO orbital parameters are taken from Ref.~\cite{mcguire_overview_2021} with apogee and perigee altitudes of 33,900 km and 350 km, respectively, and an inclination of 28.5\(\degree\). Because it is stated in Ref.~\cite{mcguire_overview_2021} that the right ascension of the ascending node (RAAN) and argument of perigee (ARGP) of the GTO are unrestricted for the initial analysis, only the specific angular momentum, eccentricity, and inclination are considered in the boundary conditions for the GTO, leaving the RAAN and ARGP as free parameters. True anomaly is also free, however, this fact is inherent to the LC law that will be used to solve the trajectory optimization problem.

The boundary conditions on the GTO are therefore defined as
\begin{align} \label{eq: GTO BCs}
    h\left(\bm{x}(t_0)\right) & = h_\text{GTO}, & e\left(\bm{x}(t_0)\right) & = e_\text{GTO}, & i\left(\bm{x}(t_0)\right) & = i_\text{GTO},
\end{align}
where subscript `\(\text{GTO}\)' denotes values of the GTO and the specific angular momentum, $h$, eccentricity, $e$, and inclination, $i$, are defined as \cite{vallado_fundamentals_2022},
\begin{align} 
    h & = \left\|\bm{r}\times\bm{v}\right\|, & e & = \left\|\frac{\left(v^2-\frac{\mu_\text{Earth}}{r}\right)\bm{r} - \left(\bm{r}^\top\bm{v}\right)\bm{v}}{\mu_\text{Earth}}\right\|, & i & = \cos^{-1}{\left(\frac{h_{\hat{\bm{z}}}}{h}\right)},
\end{align}
where \(v=\|\bm{v}\|\) and \(h_{\bm{z}}\) denotes the $z$ component of the specific angular momentum vector in the J2000 ECI frame. 

It is expected for the spacecraft's osculating orbit, with respect to the Earth, to become hyperbolic close to the Moon. Thus,  the angular momentum was selected as opposed to, for example, the semimajor axis. The semiparameter, defined as \(p=h^2/\mu_\text{Earth}\) \cite{vallado_fundamentals_2022}, could also be an acceptable element to target. Ultimately, the goal is to target the size, shape, and inclination of the GTO only. A variety of other boundary conditions could be formulated of which some could lead to a better CLF and, subsequently, a better control law and therefore will be investigated in our future work.

The NRHO ephemeris is obtained from the kernel file\footnote{\url{https://naif.jpl.nasa.gov/pub/naif/misc/MORE_PROJECTS/DSG/}} \[\verb|receding_horiz_3189_1burnApo_DiffCorr_15yr.bsp|\] described in Ref. \cite{david_e_lee_white_2019}. The boundary condition at \(t_f\) is defined as
\begin{equation} \label{eq: NRHO BC}
    \bm{x}(t_f) = \bm{x}_\text{NRHO}(t_f).
\end{equation}

The minimum-time constant-acceleration transfer trajectory optimization problem can be stated as,
\begin{subequations} \label{eq: ocp}
    \begin{align}
        \min_{\hat{\bm{\alpha}},t_0,t_f} \quad & J = t_f-t_0, \\
        \textrm{s.t.,} \quad & \text{Eqs.}~\eqref{eq: eom},\eqref{eq: control constraint},\eqref{eq: ecl constraint},\eqref{eq: time horizon},\eqref{eq: GTO BCs},\eqref{eq: NRHO BC}.
    \end{align}
\end{subequations}

A parameterized LC law based on the goal defined by Eq.~\eqref{eq: GTO BCs} will be derived and substituted into Eq.~\eqref{eq: eom}. Eq.~\eqref{eq: ocp} can then be solved as a parameter optimization problem using a heuristic algorithm in which Eq.~\eqref{eq: eom} is integrated over the time horizon given by Eq.~\eqref{eq: time horizon} with the initial condition given by Eq.~\eqref{eq: NRHO BC}. 

We note that Eq.~\eqref{eq: ecl constraint} is quite challenging to enforce directly since the number of eclipses, \(N\), is not known a priori and can also change during the iterations of the optimization process. Instead, Eq.~\eqref{eq: ecl constraint} is enforced as a soft penalty along with a penalty to further promote satisfaction of Eq.~\eqref{eq: GTO BCs} since it is not guaranteed. These two penalties and the time of flight are encoded into a single cost function that the heuristic algorithm minimizes. This cost function will be explained in detail after the control law is derived.

\section{Lyapunov Control Law}
The control-Lyapunov function (CLF) is defined as,
\begin{equation} \label{eq: control lyapunov function}
    V\left(\bm{x}\right) = \frac{1}{2}\bm{w}(\bm{x})^\top \bm{K} \bm{w}(\bm{x}),
\end{equation}
where the constraint vector, $\bm{w}(\bm{x})$, is defined as,
\begin{equation} \label{eq: constraint vector}
    \bm{w}(\bm{x}) = \begin{bmatrix} h(\bm{x}) - h_\text{GTO}, & e(\bm{x}) - e_\text{GTO}, & i(\bm{x}) - i_\text{GTO} \end{bmatrix}^\top.
\end{equation}

Note that no scaling is performed on Eq.~\eqref{eq: constraint vector} because the canonical scaling method ensures that \(h\) has the same order of magnitude of \(e\) and \(i\).  
Instead of using a diagonal parameter matrix \(\bm{K}\), a full parameter matrix is used to consider a larger family of CLFs as it is proposed in \cite{nurre2024expanding}. The \(3\times3\) positive-definite matrix \(\bm{K}\) is defined through a novel eigendecomposition method as,
\begin{equation} \label{eq: matrix K}
    \bm{K}= \bm{Q}\bm{\Lambda}\bm{Q}^\top,
\end{equation}
where the column vectors of \(\bm{Q}\) make up the eigenvectors of \(\bm{K}\) and \(\bm{\Lambda}\) is a diagonal matrix of the eigenvalues of \(\bm{K}\).  This parameterization is based on Ref.~\cite{nurre2024expanding} and allows an efficient (i.e., minimal number of parameters) representation of a full matrix that is guaranteed to be positive-definite subject only to bounds on its parameters. The eigenvalue matrix, \(\bm{\Lambda}\), is simple to construct, i.e.,
\begin{equation}
    \bm{\Lambda} = \begin{bmatrix} k_1 & 0 & 0 \\ 0 & k_2 & 0 \\ 0 & 0 & k_3 \end{bmatrix},
\end{equation}
where the parameters \(k_1\), \(k_2\), and \(k_3\) are constrained to being real and positive. The matrix \(\bm{Q}\) can be generated as a rotation matrix and Ref.~\cite{nurre2024expanding} outlines a generalized way to generate rotation matrices in \(n\)-dimensions. Because \(\bm{Q}\) is \(3\times3\), in this work, the method in Ref.~\cite{nurre2024expanding} reduces to any standard Euclidean rotation matrix parameterized by 3 angle-like parameters, \(k_4\), \(k_5\), and \(k_6\).

A control law is derived by making the time-derivative of Eq.~\eqref{eq: control lyapunov function}, \(dV/dt=\dot{V}\), as negative as possible subject to Eq.~\eqref{eq: control constraint}, i.e., we have
\begin{equation} \label{eq: argmin V dot}
    \hat{\bm{\alpha}}^* = \argmin_{\|\hat{\bm{\alpha}}\|=1}{\dot{V}}.
\end{equation}


In Eq.~\eqref{eq: argmin V dot}, \(\dot{V}\) can be found through the chain rule as,
\begin{equation} \label{eq: control lyapunov function time derivative}
    \dot{V} = \frac{\partial V}{\partial \bm{x}}\frac{\partial \bm{x}}{\partial t} + \frac{\partial V}{\partial t}.
\end{equation}

Since \(V\) does not explicitly depend on time, Eq.~\eqref{eq: control lyapunov function time derivative} reduces to 
\begin{equation}\label{eq: control lyapunov function time derivative 2}
    \dot{V} = \frac{\partial V}{\partial \bm{r}}\bm{v} + \frac{\partial V}{\partial \bm{v}}\left(\bm{a}_\text{kep} + \bm{a}_\text{3rd} + \bm{a}_{J_2} + \hat{\bm{\alpha}}a_\text{sc}\delta_\text{ecl}\right).
\end{equation}

It can be shown that Eq.~\eqref{eq: argmin V dot} is pointwise satisfied with the selection of thrust steering vector as,
\begin{equation} \label{eq: control law forwards}
    \hat{\bm{\alpha}}^* = -\left(\frac{\frac{\partial V}{\partial \bm{v}}}{\left\|\frac{\partial V}{\partial \bm{v}}\right\|}\right)^\top.
\end{equation}

Because the transfer problem is being solved backwards in time, the sign of Eq.~\eqref{eq: control law forwards} should be reversed to ensure \(V\) approaches 0 at the GTO departure time \(t_0\). The resulting control law is
\begin{equation} \label{eq: control law backwards}
    \hat{\bm{\alpha}}^* = \left(\frac{\frac{\partial V}{\partial \bm{v}}}{\left\|\frac{\partial V}{\partial \bm{v}}\right\|}\right)^\top.
\end{equation}

Eq.~\eqref{eq: control law backwards} is calculated in CasADi \cite{andersson_casadi_2019}, a symbolic framework that uses automatic differentiation. Note that if the transfer problem that departs at NRHO and arrives at GTO were solved, then Eq.~\eqref{eq: control law forwards} would be used.

Due to the very low control authority available from the low-thrust propulsion system (compared to the highly-perturbed dynamical model), the CLF time derivative, Eq.~\eqref{eq: control lyapunov function time derivative 2}, may become positive over one (or more) finite intervals, thereby not guaranteeing the system to converge as stated by the Lyapunov stability theory. However, this doesn't guarantee nonconvergence either. The results of Ref. \cite{epenoy_lyapunov-based_2019} show that converged solutions can still be found as long as the CLF time derivative is negative almost everywhere except on a finite number of small intervals. This property is also observed to be held in our numerical results.

\section{Parameter Optimization Problem}
After deriving the LC law, the trajectory optimization problem in Eq.~\eqref{eq: ocp} can now be stated as a parameter optimization problem (POP). The parameters considered are the NRHO insertion date, \(t_f\), bounded by
\begin{equation}
    t_{f,\text{lb}} \leq t_f \leq t_{f,\text{ub}},
\end{equation}
the time of flight, \(\Delta t\), bounded by 
\begin{equation}
    \Delta t_\text{lb} \leq \Delta t \leq \Delta t_\text{ub},
\end{equation}
and finally the 6 CLF parameters in Eq.~\eqref{eq: matrix K}, bounded by
\begin{align}
    & 0 < k_i \leq k_\text{ub},~\text{for}~i=1,2,3, & & 0 \leq k_4 \leq \pi, & & 0 \leq k_j \leq 2\pi,~\text{for}~j=5,6.
\end{align}

These parameters are optimized using MATLAB's particle swarm optimization (PSO), a stochastic optimization algorithm that optimizes a scalar cost function subject only to bounds on the design variables. Under this parametrization, the time horizon in Eq.~\eqref{eq: time horizon} can be expressed as,
\begin{equation} \label{eq: new time horizon}
    t \in [t_f,t_f - \Delta t].
\end{equation}

An important step in the resulting POP is to accurately solve for the initial value problem (IVP) given by the set of ordinary differential equation (ODEs) in Eq.~\eqref{eq: eom} with the control law in Eq.~\eqref{eq: control law backwards} and boundary condition given by Eq.~\eqref{eq: NRHO BC} over the time horizon in Eq.~\eqref{eq: new time horizon}, i.e.,
\begin{align} \label{eq: ode}
    \dot{\bm{x}} & = \bm{f}\left(t,\bm{x},\hat{\bm{\alpha}}^*;\bm{K}\right), & & \bm{x}(t_f) = \bm{x}_\text{NRHO}(t_f), & t & \in[t_f,t_f-\Delta t].
\end{align}

MATLAB's variable-step variable-order nonstiff ODE integrator \verb|ode113| is used with an absolute and relative tolerance of \(1.0\times10^{-10}\). Our extensive numerical studies indicate that this integrator performed most efficiently with the prescribed tolerances against the rest of MATLAB's ODE integrators.

The event-detection capability of \verb|ode113| is used extensively while solving Eq.~\eqref{eq: ode}. The method works by monitoring the sign of an \(M\) number of scalar event functions, \(e_m\) for \(m=1,\dots,M\). When the \(m\)-th function changes sign, a regula falsi algorithm is used to find the precise location of the zero of \(e_m\), and, if all the corresponding termination conditions are met, integration stops. In our paper, there are $M=7$ event functions defined as follows,
\begin{subequations} \label{eq: event functions}
    \begin{align}
        e_1 & = r - R_\text{Earth} - 200~\text{km}, \label{eq: e1} \\
        e_2 & = r - R_\text{Moon} - 200~\text{km}, \label{eq: e2} \\
        e_3 & = |h - h_\text{GTO}| - \epsilon, \label{eq: e3} \\
        e_4 & = |e - e_\text{GTO}| - \epsilon, \label{eq: e4} \\
        e_5 & = |i - i_\text{GTO}| - \epsilon, \label{eq: e5} \\
        e_6 & = S_\text{ecl,Earth,1}, \label{eq: e6} \\
        e_7 & = S_\text{ecl,Moon,1}. \label{eq: e7}
    \end{align}
\end{subequations}

Eqs.~\eqref{eq: e1} and \eqref{eq: e2} monitor if the spacecraft has intersected 200 km above the surface of the Earth and Moon. If either of their signs change, then integration stops and the cost function is appropriately updated and returned to PSO. Eqs.~\eqref{eq: e3}, \eqref{eq: e4}, and \eqref{eq: e5} monitor if the solution has converged or not (i.e., orbit insertion has been achieved or not). If any one of their signs become negative while the other two are also negative, then integration stops and the cost function is appropriately updated and returned to PSO. The tolerance \(\epsilon=1.0\times10^{-3}\) was  chosen as it provides a balance between convergence speed and accuracy; however, future work should investigate using different convergence criteria. 

Eqs.~\eqref{eq: e6} and \eqref{eq: e7} are from Eqs.~\eqref{eq: earth ecl switching functions} and \eqref{eq: moon ecl switching functions}, respectively, and determine if the spacecraft is in an eclipse or not. If Eq.~\eqref{eq: earth ecl switching functions} (resp. Eq.~\eqref{eq: moon ecl switching functions}) changes sign while \(S_\text{ecl,Earth,2}\) (resp. \(S_\text{ecl,Moon,2}\)) is negative, then integration is terminated. However, integration is then restarted from the same time and state. This logic is followed so that the discrete function in Eq.~\eqref{eq: ecl function} is modeled as accurately as possible.

In formulating the cost function, let \(t_\text{end}\) denote the final time returned by Eq.~\eqref{eq: ode} under the event-detection logic, i.e., \(t_\text{end}\) always satisfies \(t_f-\Delta t\leq t_\text{end}\leq t_f\). The first priority of the cost function is to ensure the solution converges. If a solution to Eq.~\eqref{eq: ode} does not satisfy the constraint below,
\begin{equation} \label{eq: convergence criteria}
    \left|\bm{w}(\bm{x}(t_\text{end}))\right|<\epsilon,
\end{equation}
then, the value of cost, $J_1$, defined as,
\begin{equation} \label{eq: convergence cost}
    J_1=\|\bm{w}(\bm{x}(t_\text{end}))\|,
\end{equation}
is returned. Because of the highly nonlinear dynamics in the vicinity of the Moon, it was found that solutions commonly intersect the surface of the Moon. Thus, if integration was stopped due to Eq.~\eqref{eq: e2} becoming negative, then, 
\begin{equation} \label{eq: moon crash cost}
    J_2 = 1000\|\bm{w}(\bm{x}(t_\text{end}))\|,
\end{equation}
is returned as the cost function to PSO as a penalization.

If Eq.~\eqref{eq: convergence criteria} is satisfied for a solution, but Eq.~\eqref{eq: ecl constraint} is not, then 
\begin{equation} \label{eq: ecl penalty cost}
    J_3 = -\left(\sum_{n = 1}^N \max{\left(d_{\text{ecl},n}-90~\text{minutes},0\right)}\right)^{-1},
\end{equation}
is returned as the cost function to PSO. Note that it appears there's a possibility for a division by zero in Eq.~\eqref{eq: ecl penalty cost}, however, the expression in Eq.~\eqref{eq: ecl penalty cost} is not evaluated if Eq.~\eqref{eq: ecl constraint} is satisfied, which eliminates this possibility. Also, Eq.~\eqref{eq: ecl penalty cost} is made negative to differentiate it from Eqs.~\eqref{eq: convergence cost} and \eqref{eq: moon crash cost}, but inverted so that the violated eclipse durations are still minimized. 

Finally, if Eqs.~\eqref{eq: convergence criteria} and \eqref{eq: ecl constraint} are satisfied for a solution, then 
\begin{equation} \label{eq: tf cost}
    J_4 = -\left(t_f-t_\text{end}\right)^{-1},
\end{equation}
is returned as the cost function to PSO. This cost is also made negative to differentiate from Eqs.~\eqref{eq: convergence cost} and \eqref{eq: moon crash cost} and inverted so that time of flight is minimized. However, to ensure it is differentiated from Eq.~\eqref{eq: ecl penalty cost}, it has units of \(1/[\text{years}]\) while Eq.~\eqref{eq: ecl penalty cost} has units of \(1/[\text{seconds}]\) so that they are on different orders of magnitude. Further, to ensure that minuscule eclipse violations aren't interpreted as extremely low times of flights, if \(J_3<J_4\) occurs for a converged solution, then Eq.~\eqref{eq: tf cost} is returned instead of Eq.~\eqref{eq: ecl penalty cost}. While this allows for solutions with eclipses longer than 90 minutes to be deemed feasible, these solutions will only be infeasible by a duration on the order of a second. Note that the cost function used in this work is not continuous due to the logic involved, however, stochastic optimization algorithms, such as PSO, can deal with discontinuous cost functions. 

Because LC laws are prone to extreme oscillations/chattering at the end of the maneuver, the step size of variable step integrators can become minuscule and halt progress \cite{noble_ariel_hatten_critical_2012}. To overcome this issue, integration of \verb|ode113| is stopped when a certain number of function evaluations is reached. A simple logic is implemented inside \verb|ode113| and, if triggered, then Eq.~\eqref{eq: convergence cost} is simply returned as the cost function to PSO. In this paper, \(1.0\times10^{5}\) function evaluations were arbitrarily chosen and found to provide acceptable results; however, different values may further benefit the algorithm given that the number of iterations is a problem-dependent number. The event-detection logic and cost function values are summarized in Algorithm \ref{alg: event detection}.

\begin{algorithm}[]
\caption{Event-detection and cost function logic}\label{alg: event detection}
\begin{algorithmic}
\For{Every integration step while integrating Eq.~\eqref{eq: ode}} \Comment{Numerical integration of IVP}
\If{\(e_1<0\)} \Comment{Check if spacecraft hits Earth}
    \State Terminate integration
    \State Return \(J_1\) as cost to PSO \Comment{See Eq.~\eqref{eq: convergence cost}}
\ElsIf{\(e_2<0\)} \Comment{Check if spacecraft hits Moon}
    \State Terminate integration
    \State Return \(J_2\) as cost to PSO \Comment{See Eq.~\eqref{eq: moon crash cost}}
\ElsIf{\(e_i<0\) for any \(i\in\{3,4,5\}\)}
    \If{\(e_j<0\) for all \(j=\{3,4,5\}\setminus i\)} \Comment{Check if solution converges}
        \State Terminate integration
        \If{Eq.~\eqref{eq: ecl constraint} is satisfied} \Comment{Check if eclipse constraint is satisfied}
            \State Return \(J_4\) as cost to PSO \Comment{See Eq.~\eqref{eq: tf cost}}
        \ElsIf{\(J_3<J_4\)} \Comment{Ensure Eq.~\eqref{eq: ecl penalty cost} is not less than Eq.~\eqref{eq: tf cost} due to extremely small eclipse violations}
            \State Return \(J_4\) as cost to PSO \Comment{See Eq.~\eqref{eq: tf cost}}
        \Else
            \State Return \(J_3\) as cost to PSO \Comment{See Eq.~\eqref{eq: ecl penalty cost}}
        \EndIf
    \EndIf
\ElsIf{\(e_6<0\)}
    \If{\(S_\text{ecl,Earth,2}<0\)} \Comment{Check for Earth eclipses}
        \State Terminate integration and restart from same state and time
    \EndIf
\ElsIf{\(e_7<0\)}
    \If{\(S_\text{ecl,Moon,2}<0\)} \Comment{Check for Moon Eclipses}
        \State Terminate integration and restart from same state and time
    \EndIf
\ElsIf{Number of function evaluations exceed the limit}
    \State Terminate integration
    \State Return \(J_1\) as cost to PSO \Comment{See Eq.~\eqref{eq: convergence cost}}
\Else
    \State Return \(J_1\) as cost to PSO \Comment{See Eq.~\eqref{eq: convergence cost}}
\EndIf
\EndFor
\end{algorithmic}
\end{algorithm}

\section{Results}
The results are presented for a transfer problem  with a fixed NRHO insertion date and then with a variable NRHO insertion date. These two cases are considered to demonstrate the impact of the insertion date and the types of solutions that can be achieved. The fixed NRHO insertion date was arbitrarily selected as \(t_f = \) 2026 DEC 06 00:00:00 UTC. This coincides with a state that is roughly at apolune on the NRHO. When \(t_f\) is allowed to vary, it is bounded between \(t_{f,\text{lb}}=\) 2026 NOV 06 00:00:00 UTC and \(t_{f,\text{ub}}=\) 2027 JAN 05 00:00:00 UTC, or, \(t_{f,\text{lb}}=t_f - 30\) days and \(t_{f,\text{ub}}=t_f+30\) days. The value of 30 days was selected because it is approximately equal to 1 period of the Moon's orbit around the Earth, giving a variety of phasing possibilities to the solution space. The time of flight for all cases was bounded by \(\Delta t_\text{lb}=200\) days and \(\Delta t_\text{ub}=400\) days. Simulations were performed on a 2023 MacBook Pro with the Apple M2 Pro chip, which allows for 10 ``workers'' in MATLAB to run PSO in parallel.

\subsection{Fixed NRHO Insertion Date}
\begin{figure}
    \centering \includegraphics[width=0.85\textwidth]{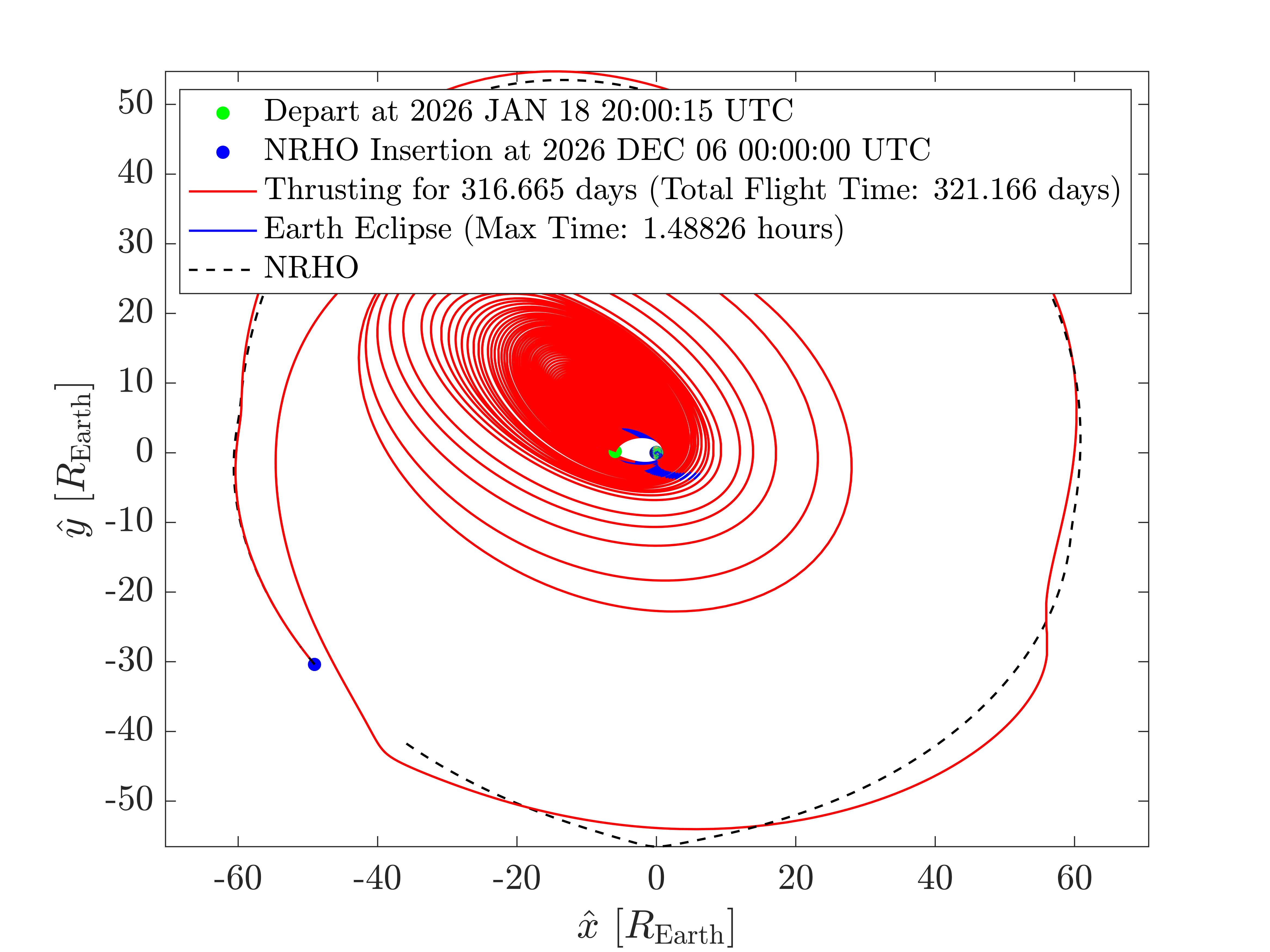} \\
    \centering \includegraphics[width=0.85\textwidth]{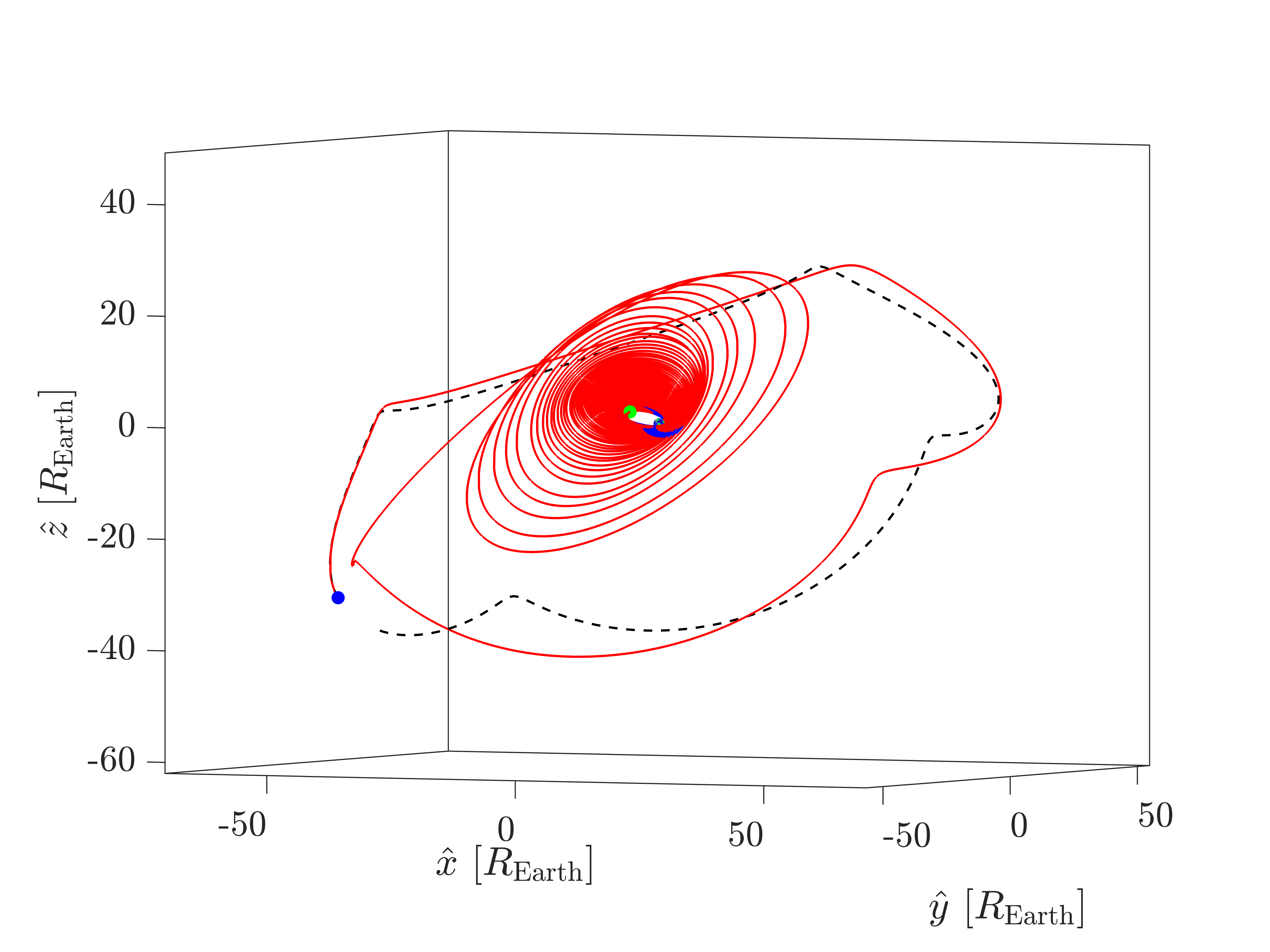}
    \caption{Trajectory in J2000 ECI frame for transfer solution with a fixed NRHO insertion date.} 
    \label{fig: traj eci fixed}
\end{figure}

\begin{figure}
    \centering \subfloat[Earth-centered Sun-Earth rotating frame.]{\includegraphics[width=0.85\textwidth]{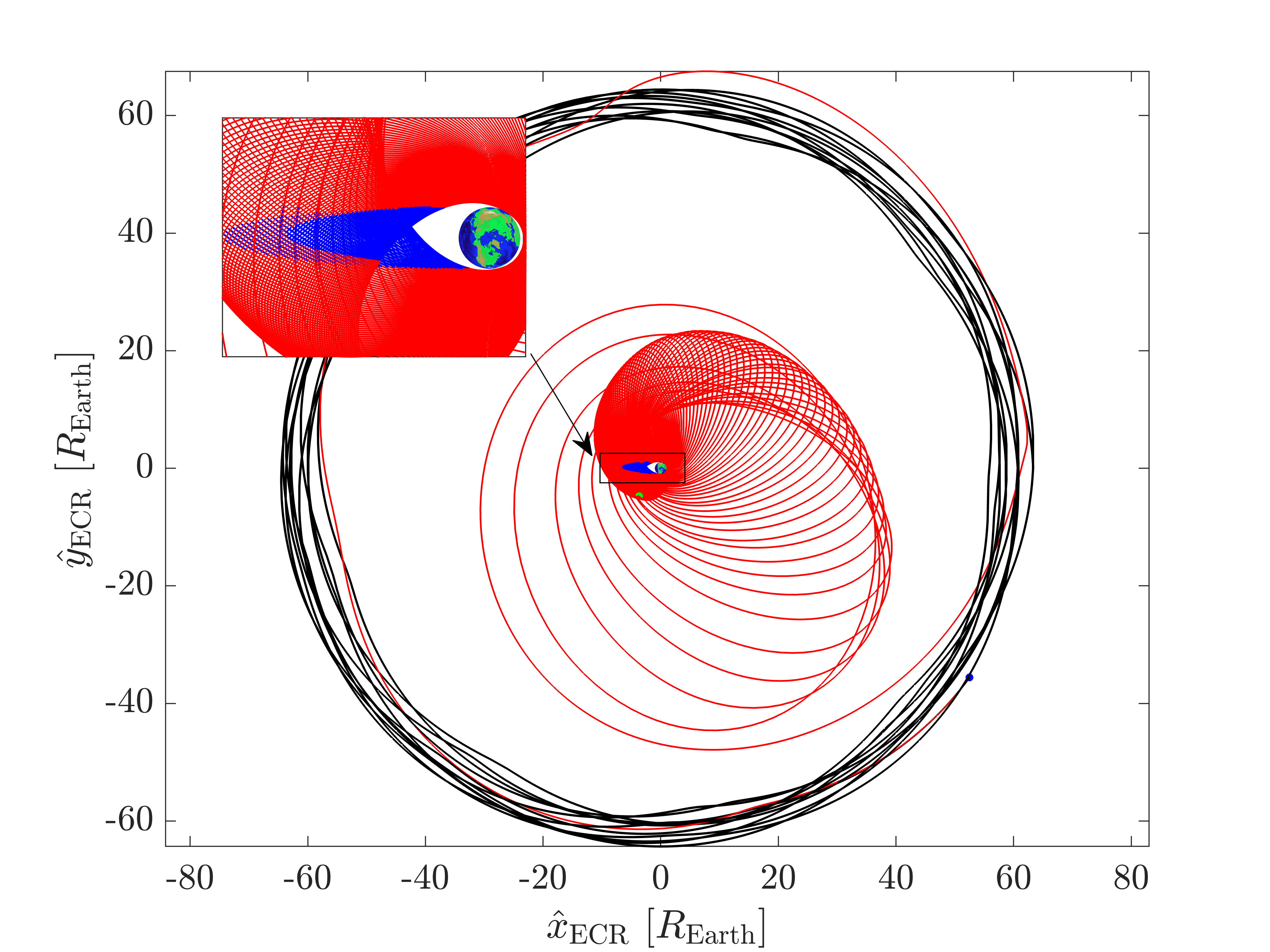} \label{fig: traj ecr fixed}}
    \\
    \centering \subfloat[Moon-centered Earth-Moon rotating frame]{\includegraphics[width=0.85\textwidth]{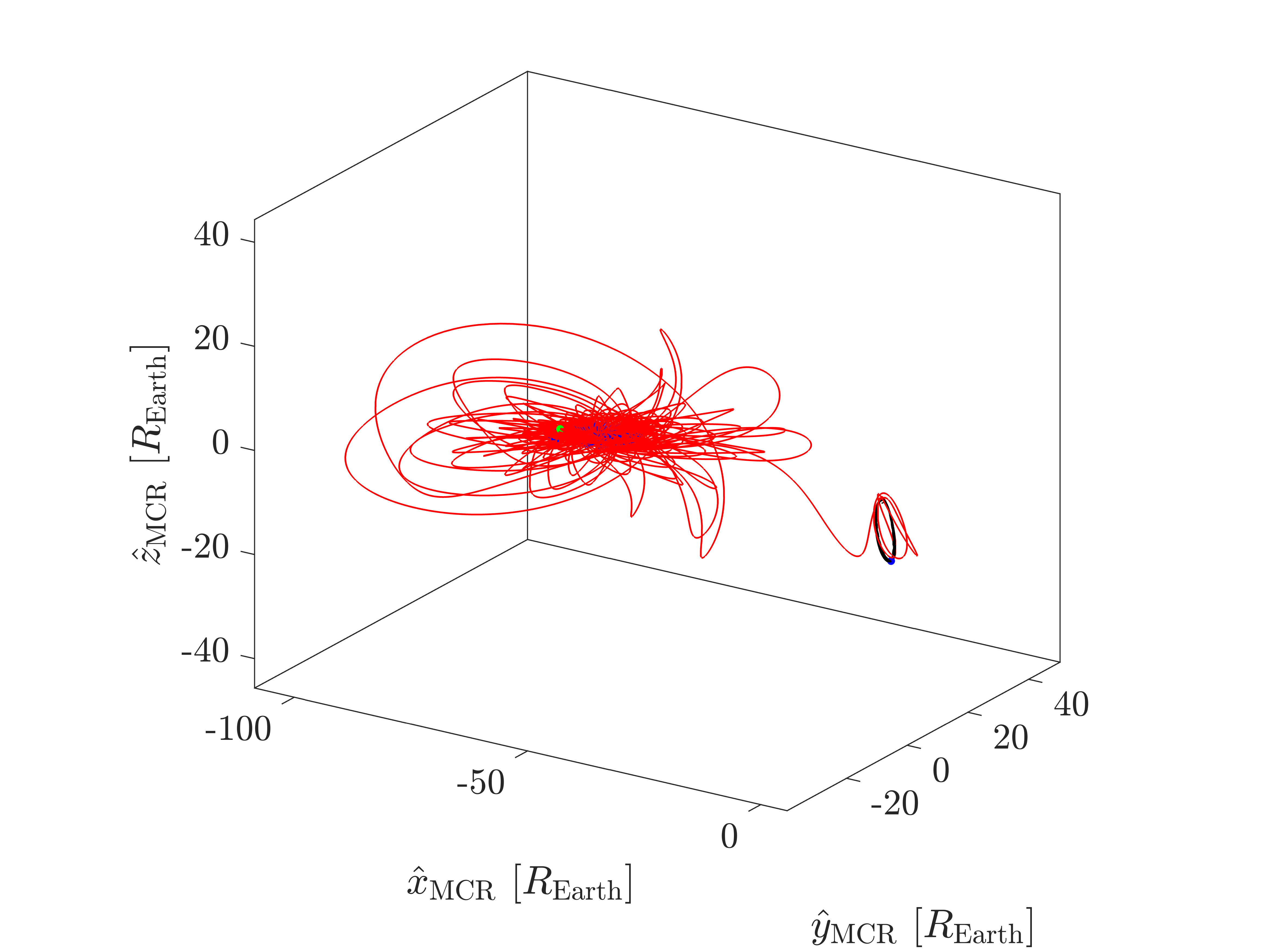} \label{fig: traj mcr fixed}} 
    \caption{Trajectories for transfer solution with a fixed NRHO insertion date.} 
    \label{fig: traj 2}
\end{figure}

For this transfer problem, PSO was run 5 times with a swarm size of 500 and a maximum time limit of 1 hour. The best run provided an eclipse-feasible solution, with a time of flight of \(\Delta t=\) 321.17 days. The trajectory for this solution, in the J2000 ECI frame, is shown in Figure \ref{fig: traj eci fixed}. The trajectory is also shown in the Earth-centered Sun-Earth rotating frame in Figure \ref{fig: traj ecr fixed} and in the Moon-centered Earth-Moon rotating frame in Figure \ref{fig: traj mcr fixed}. These frames are denoted by the unit vectors \(\left\{\hat{\bm{x}}_\text{ECR},\,\hat{\bm{y}}_\text{ECR},\,\hat{\bm{z}}_\text{ECR}\right\}\) and \(\left\{\hat{\bm{x}}_\text{MCR},\,\hat{\bm{y}}_\text{MCR},\,\hat{\bm{z}}_\text{MCR}\right\}\), respectively, where the subscript `ECR' denotes Earth-centered rotating and `MCR' denotes Moon-centered rotating. Note that the legend in Figure \ref{fig: traj eci fixed} also applies to Figure \ref{fig: traj ecr fixed} and Figure \ref{fig: traj mcr fixed}.

\begin{figure}
    \centering \subfloat[Specific angular momentum.]{\includegraphics[width=0.55\textwidth]{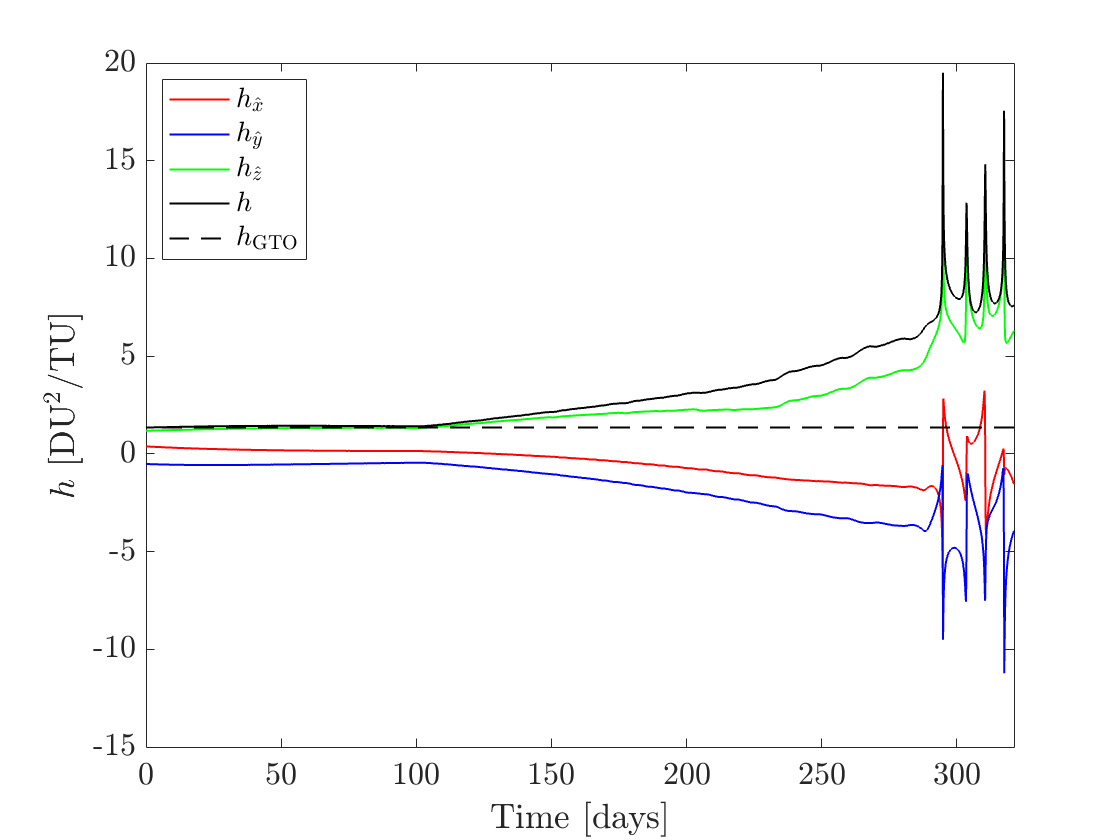} \label{fig: h fixed}}
    \\
    \centering \subfloat[Eccentricity.]{\includegraphics[width=0.55\textwidth]{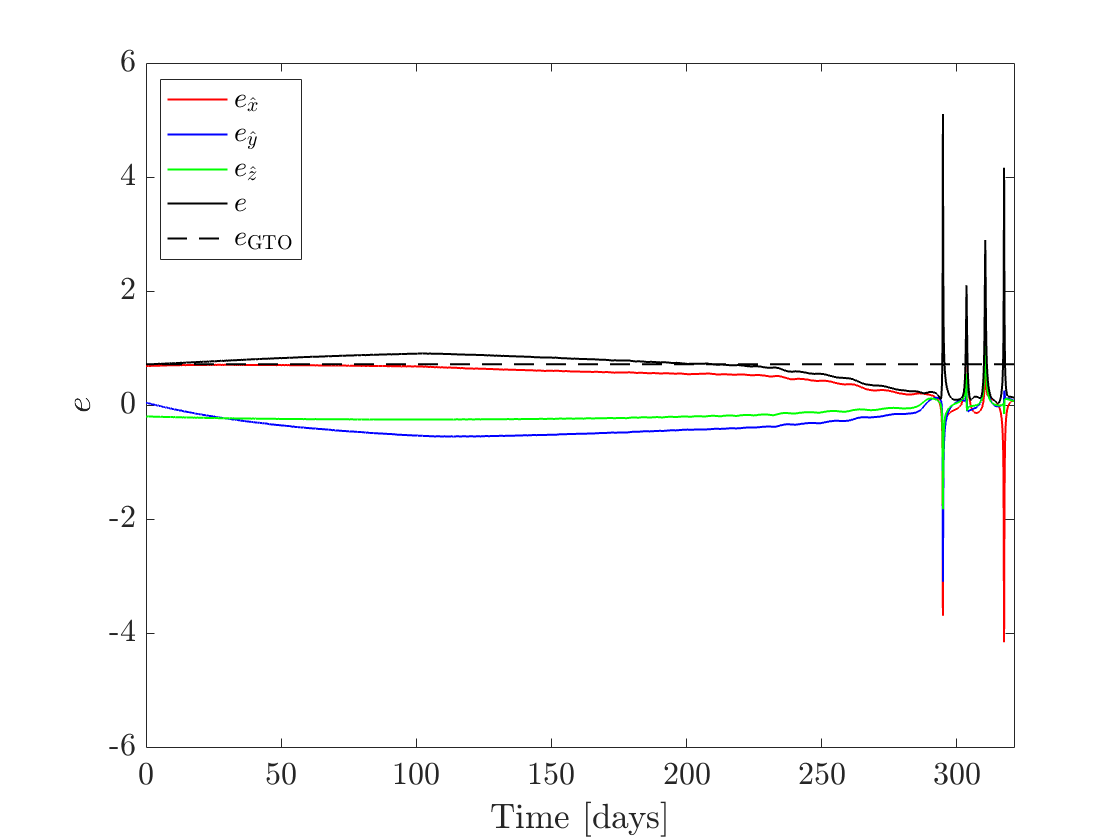} \label{fig: e fixed}}
    \\
    \centering \subfloat[Inclination.]{\includegraphics[width=0.55\textwidth]{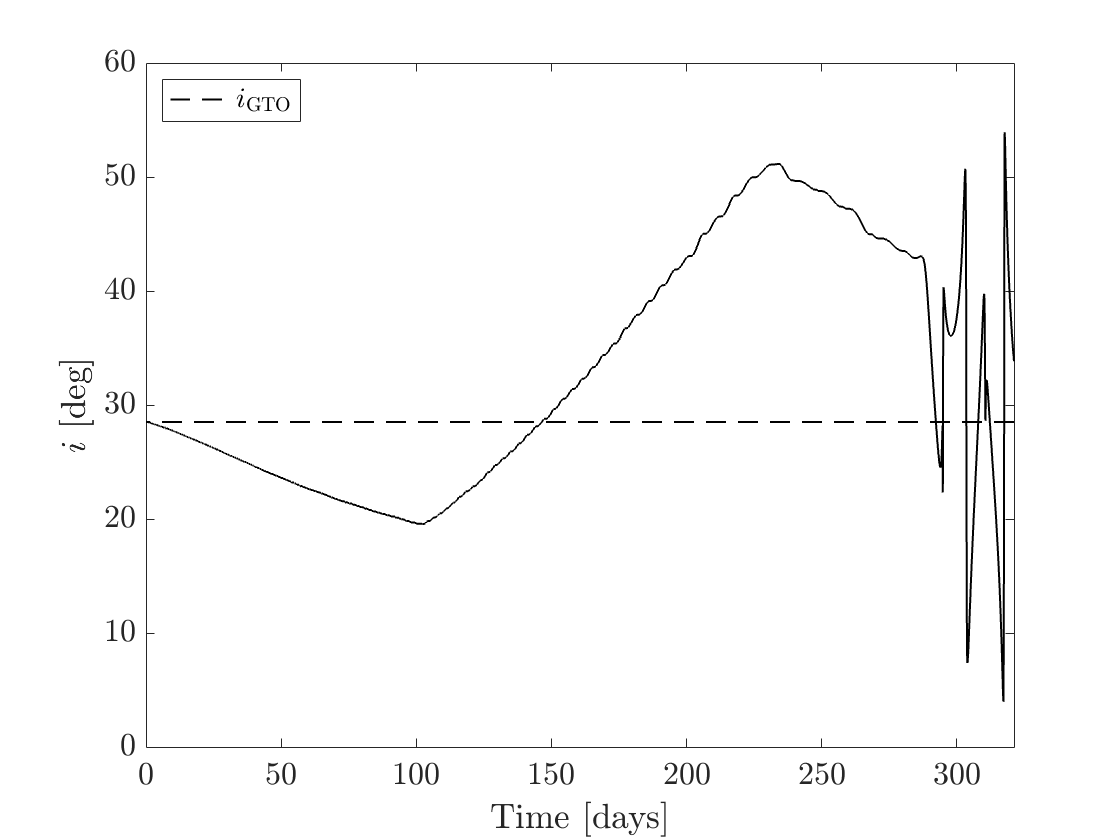} \label{fig: i fixed}}
    \caption{Orbital element time histories for the transfer solution with a fixed NRHO insertion date.} 
    \label{fig: elements fixed}
\end{figure}

\begin{figure}
    \centering \subfloat[Lyapunov function (Eq.~\eqref{eq: control lyapunov function}).] {\includegraphics[width=0.85\textwidth]{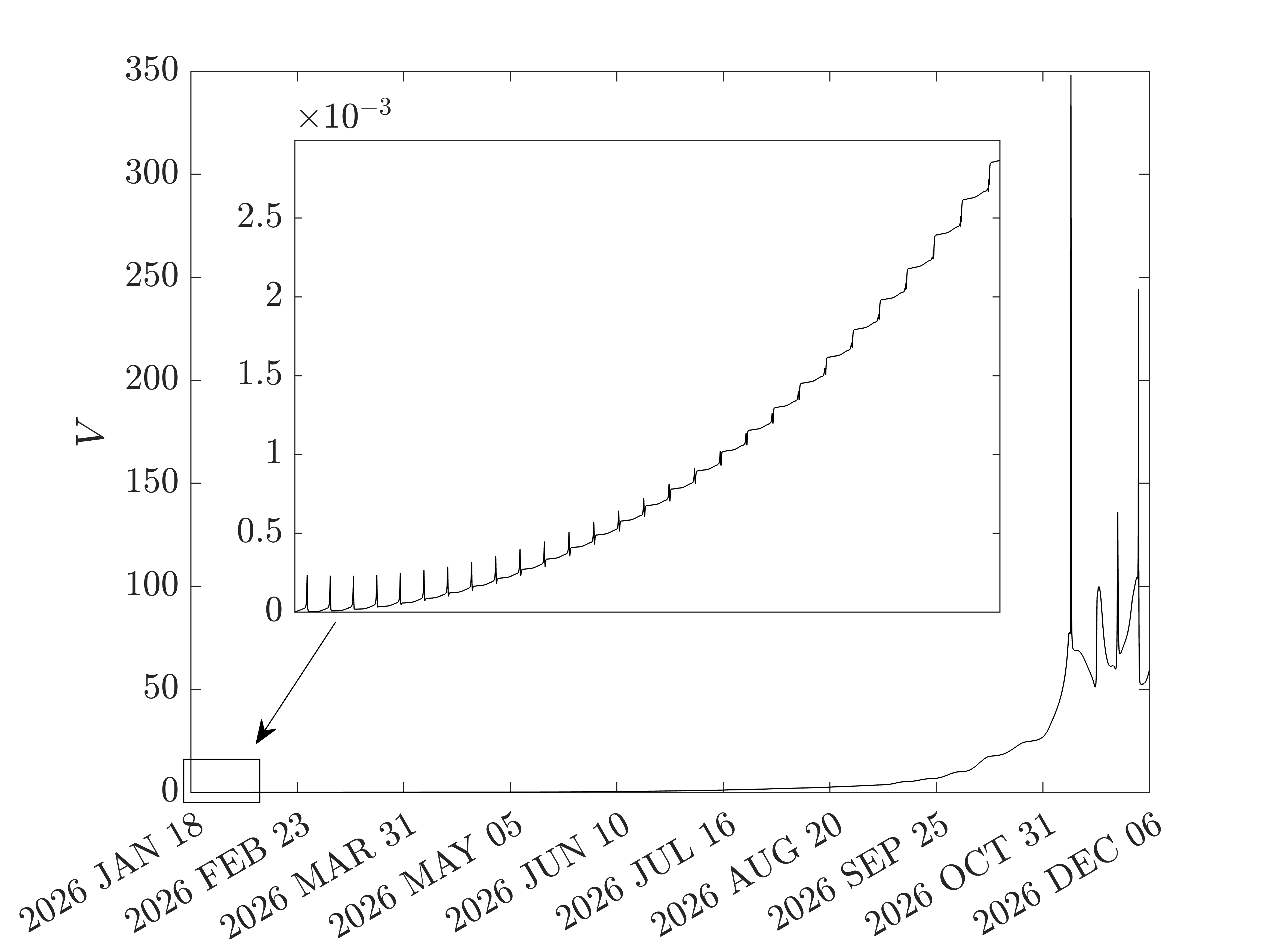} \label{fig: V fixed}}
    \\
    \centering \subfloat[Lyapunov function time derivative (Eq.~\eqref{eq: control lyapunov function time derivative 2}).]{\includegraphics[width=0.85\textwidth]{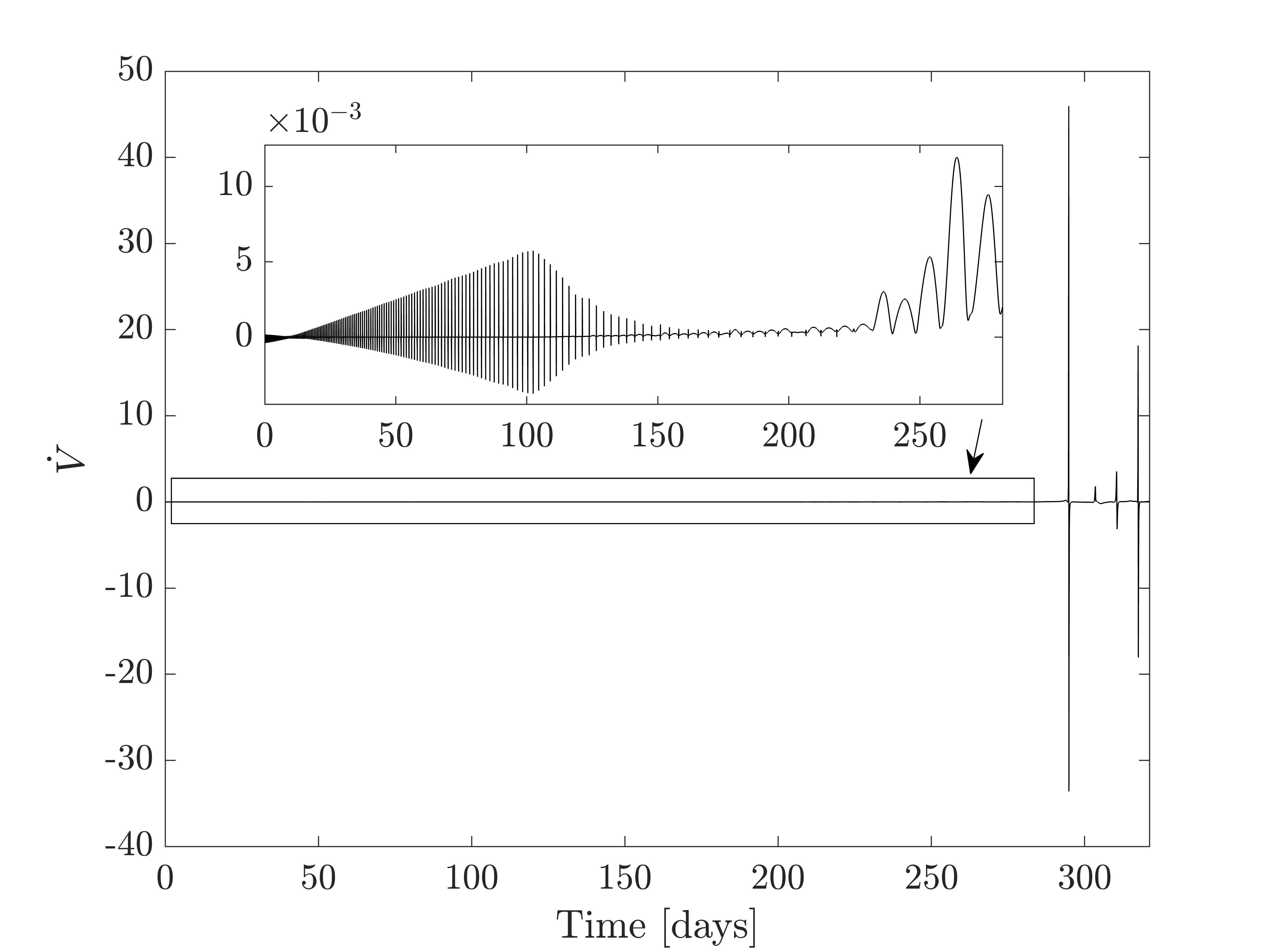} \label{fig: V dot fixed}}
    \caption{Lyapunov function time histories for the solution with a fixed NRHO insertion date.} 
    \label{fig: lyapunov}
\end{figure}

Figures \ref{fig: h fixed}, \ref{fig: e fixed}, and \ref{fig: i fixed} show the time histories of the specific angular momentum, eccentricity, and inclination, respectively. The time histories are plotted in the ``forward sense of time,'' i.e., the \(x\)-axis is 0 when the spacecraft is at the GTO and is \(\Delta t\) when the spacecraft is at the NRHO. Figures \ref{fig: V fixed} and \ref{fig: V dot fixed} show the CLF value (Eq.~\eqref{eq: control lyapunov function}) and the CLF time derivative value (Eq.~\eqref{eq: control lyapunov function time derivative 2}), respectively, and are also plotted in the forward sense of time. Note that because this problem is being solved backwards in time, the sign of the control law was reversed, so, ideally, the function in Fig. \ref{fig: V dot fixed} is positive definite. It can be observed though that the CLF time derivative, in fact, becomes negative over multiple intervals. However, the trajectory still converges to the target orbit.

\begin{figure}
    \centering \subfloat[Eclipse-triggered throttle factor (Eq.~\eqref{eq: ecl function}).]{\includegraphics[width=0.85\textwidth]{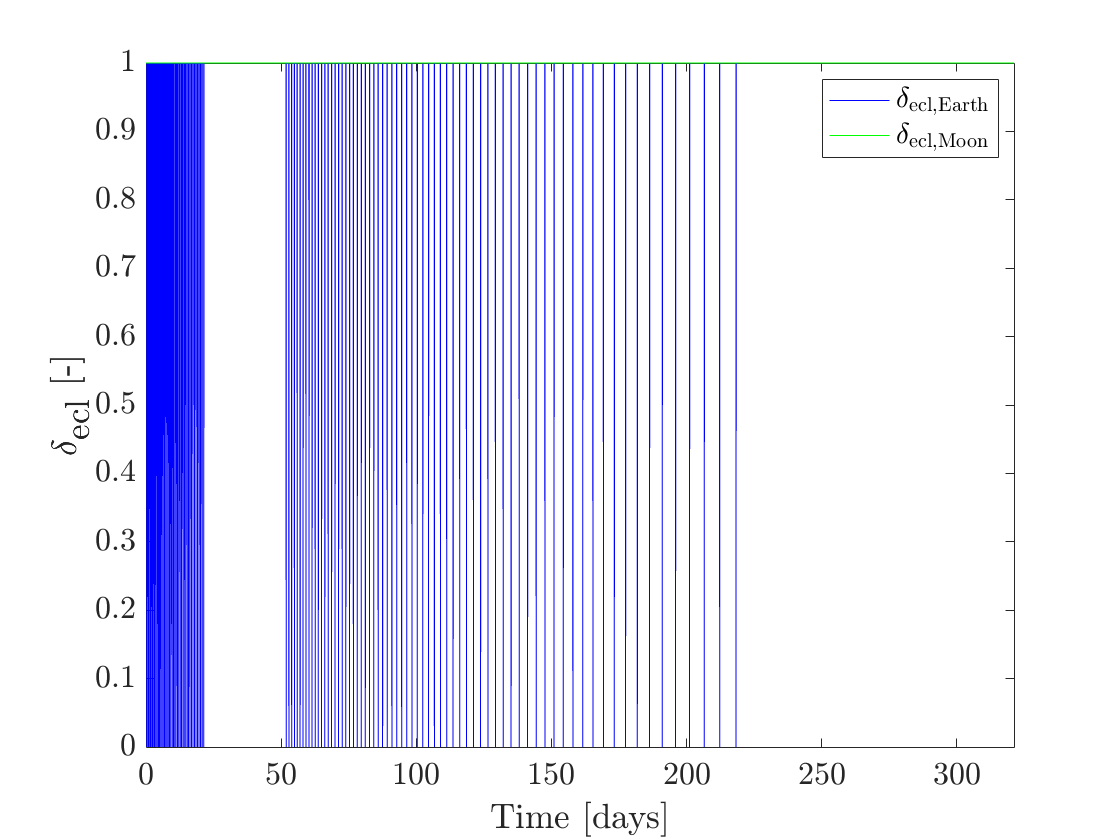} \label{fig: ecl delta fixed}}
    \\
    \centering \subfloat[Eclipse switching functions (Eqs.~\eqref{eq: earth ecl switching functions} and \eqref{eq: moon ecl switching functions}).]{\includegraphics[width=0.85\textwidth]{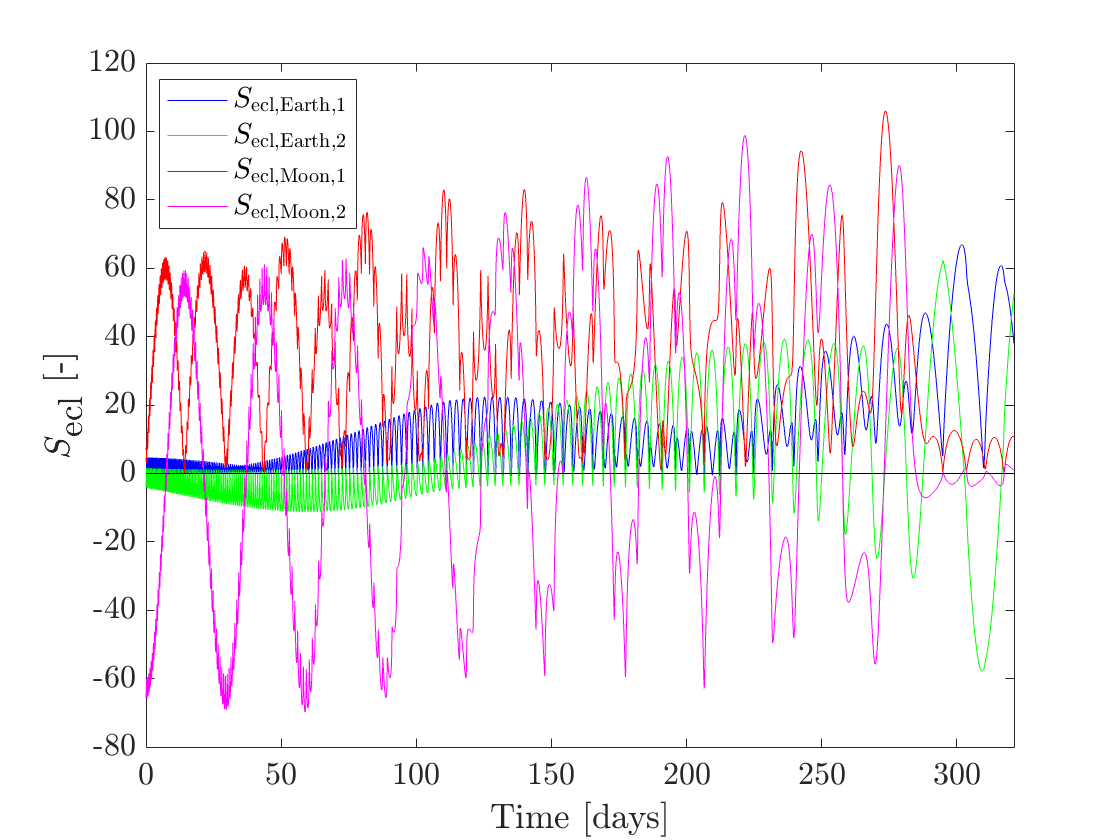} \label{fig: ecl sf fixed}}
    \caption{Eclipse function time histories for the solution with a fixed NRHO insertion date.} 
    \label{fig: eclipse functions}
\end{figure}

\begin{figure}
    \centering \subfloat[Duration of each eclipse for the best solution with a fixed NRHO insertion date.]{\includegraphics[width=0.85\textwidth]{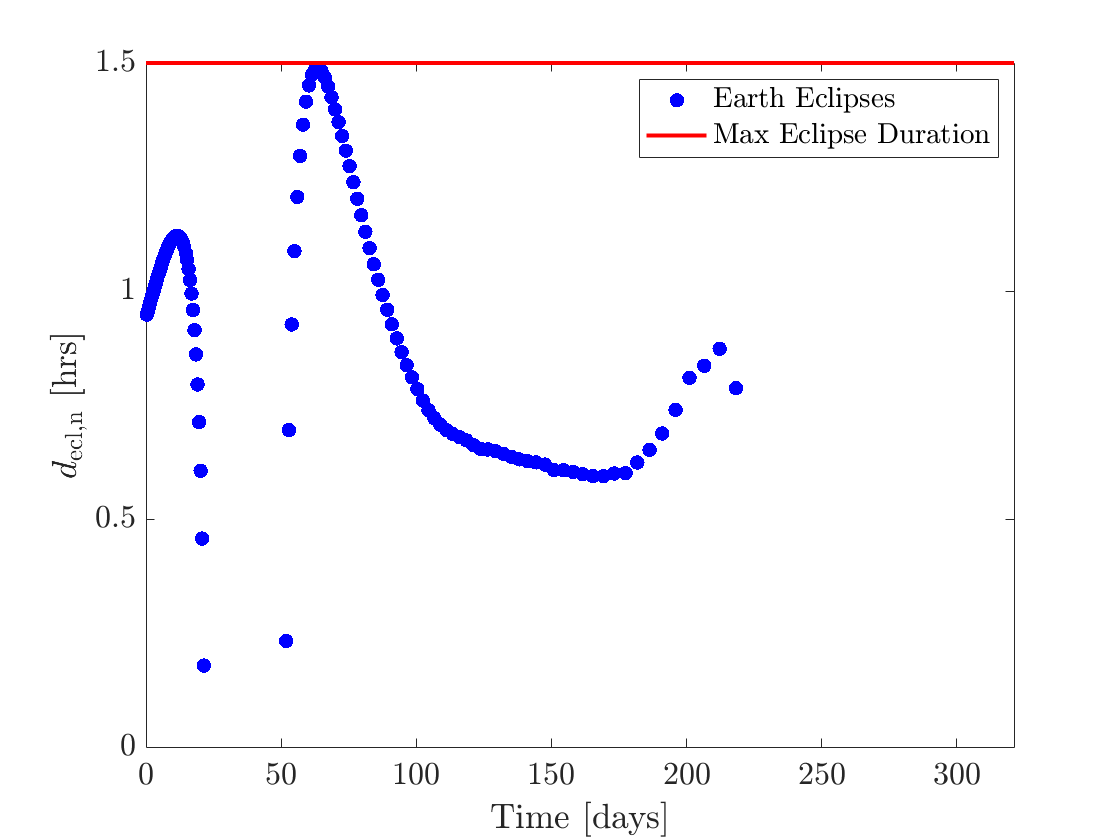} \label{fig: ecl duration fixed}}
    \\
    \centering \subfloat[Duration of each eclipse for the first feasible solution with a fixed NRHO insertion date.]{\includegraphics[width=0.85\textwidth]{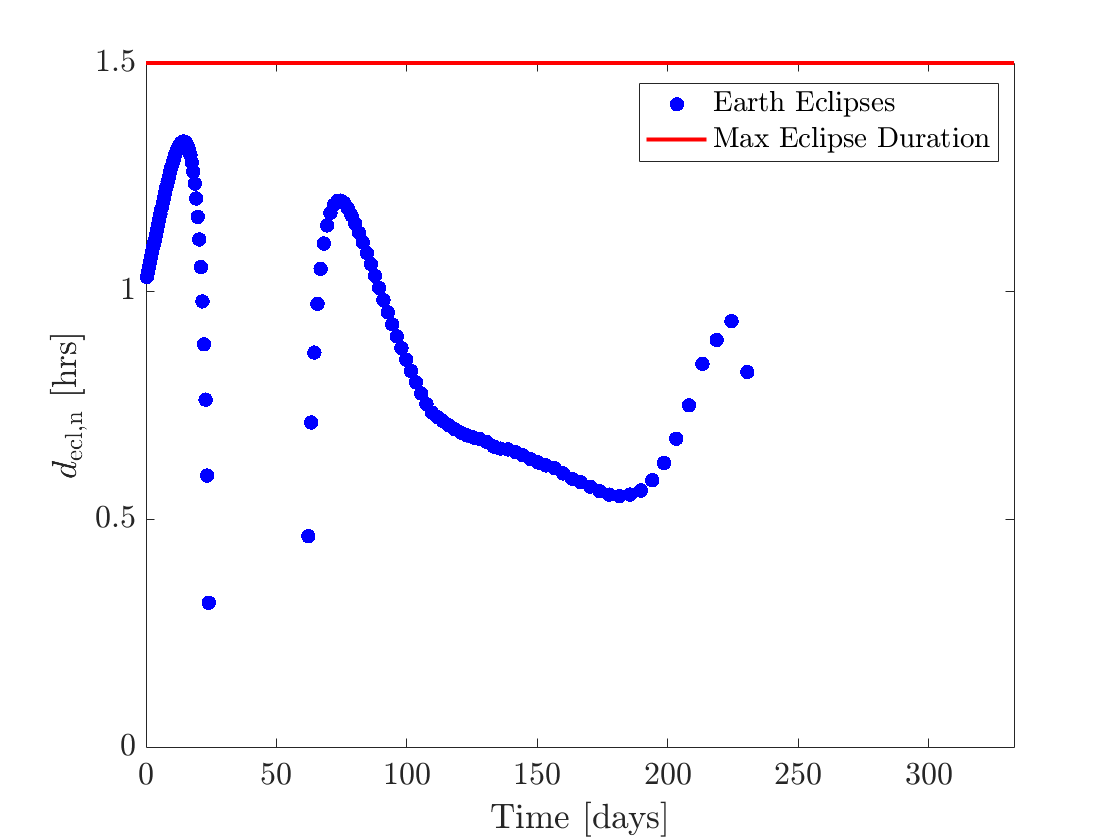} \label{fig: ecl duration fixed first}}
    \caption{Eclipse durations for two different solutions with a fixed NRHO insertion date.} 
    \label{fig: eclipse durations fixed}
\end{figure}

Figures \ref{fig: ecl delta fixed} and \ref{fig: ecl sf fixed} show the eclipse-triggered throttle factor and eclipse switching functions for the solution. Figure \ref{fig: ecl duration fixed} shows the duration of each eclipse. While included in the model, no eclipses due to the Moon occur in this solution. The first feasible solution from PSO in the same run was obtained in about 15 minutes and had a time of flight of \(\Delta t=\) 332.72 days. Figure \ref{fig: ecl duration fixed first} shows the duration of each of its eclipses. Comparing Figure \ref{fig: ecl duration fixed} and Figure \ref{fig: ecl duration fixed first}, it can be interpreted for this particular case that PSO improves the time of flight by increasing the eclipse durations as much as possible. 

\subsection{Variable NRHO Insertion Date}
\begin{figure}
    \centering \subfloat[J2000 ECI frame.]{\includegraphics[width=0.85\textwidth]{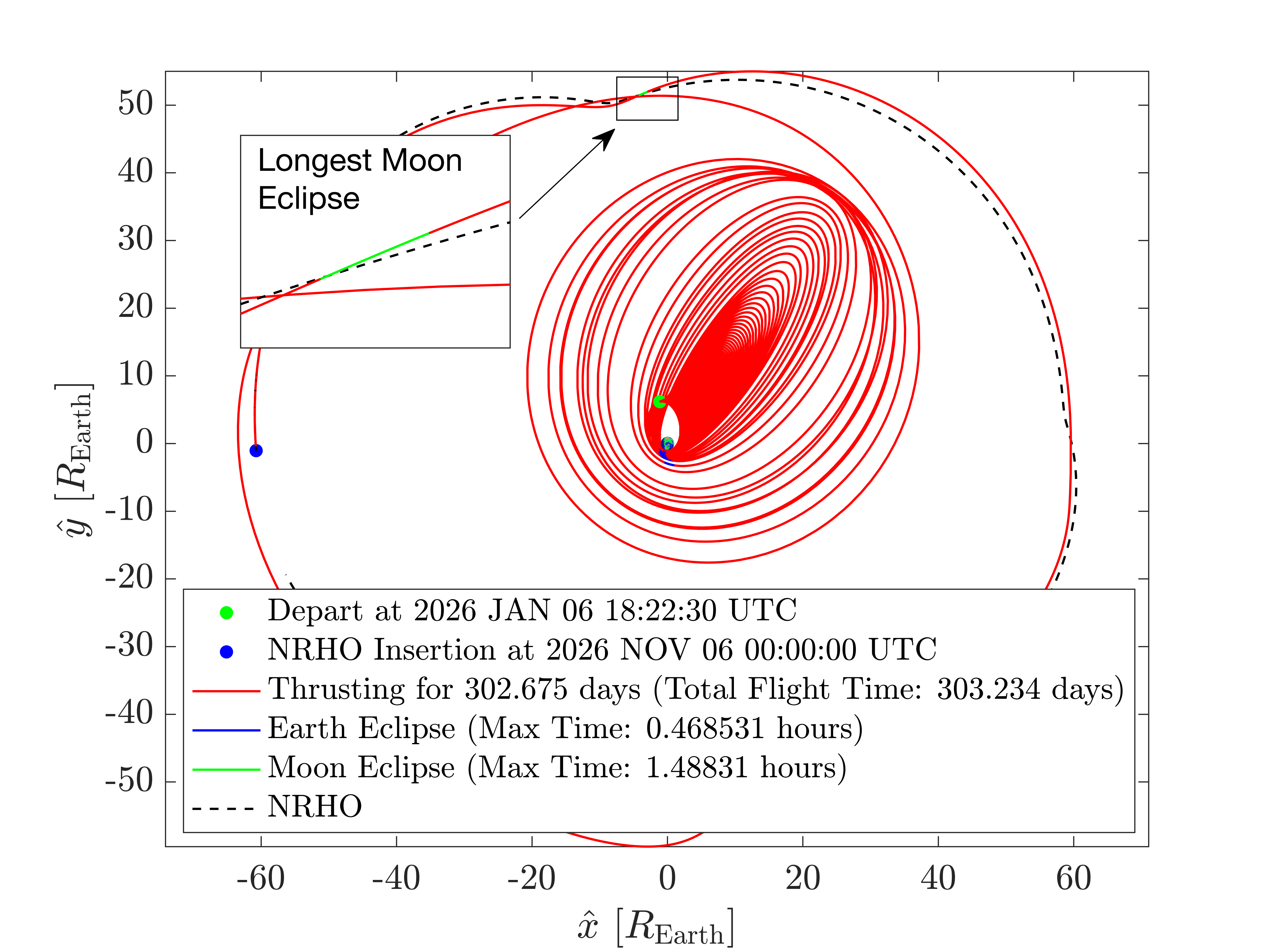} \label{fig: traj eci variable}}
    \\
    \centering \subfloat[Duration of each eclipse.]{\includegraphics[width=0.85\textwidth]{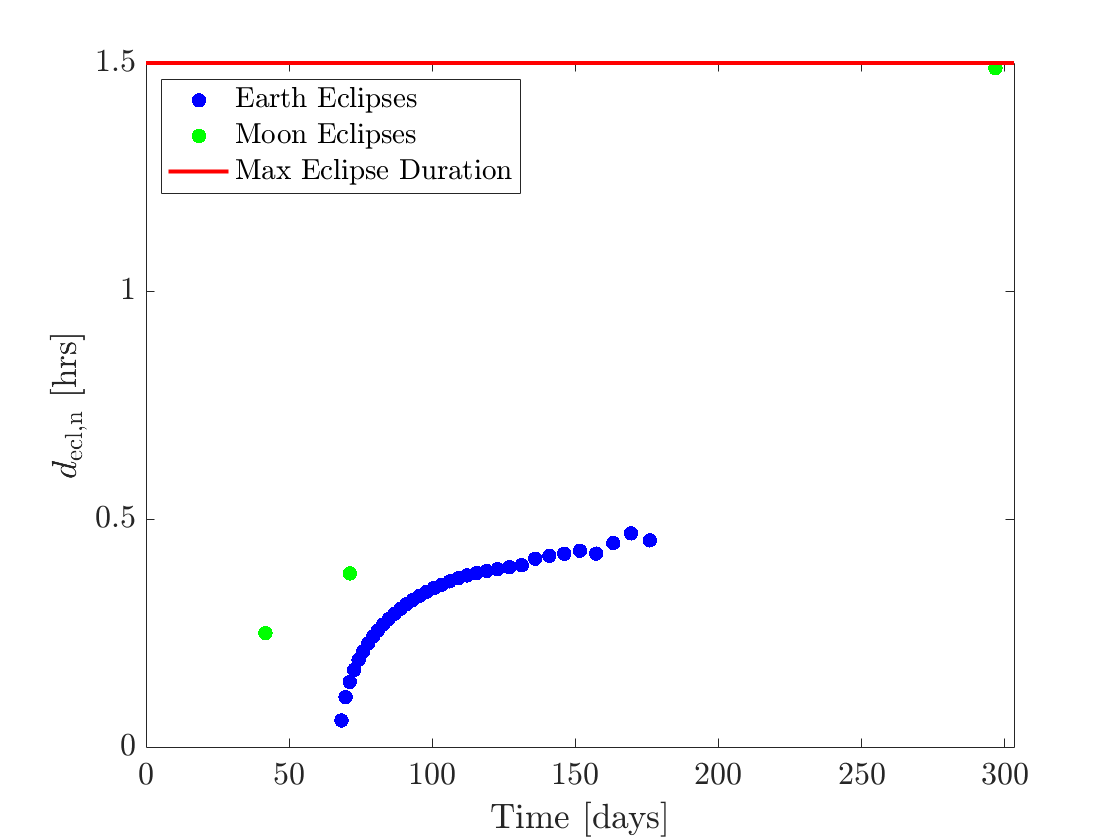} \label{fig: ecl duration variable}}
    \caption{Eclipse durations for the best solution with a variable NRHO insertion date.} 
    \label{fig: plots variable}
\end{figure}

For this transfer problem, PSO was ran once with an increased swarm size of 1000 to account for the extra parameter, \(t_f\), being optimized. The first eclipse-feasible solution was obtained after about 1 hour and 45 minutes. The NRHO insertion date is \(t_f=\) 2026 NOV 06 00:01:35 UTC with a time of flight of \(\Delta t=\) 314.03 days. The best eclipse-feasible solution was obtained after about 3 hours and 30 minutes and had an NRHO insertion date of \(t_f=\) 2026 NOV 06 00:00:00 UTC and a time of flight of \(\Delta t=\) 303.23 days. The NRHO insertion date of many of the eclipse-feasible solutions tends towards \(t_{f,\text{lb}}\), suggesting that shifting \(t_{f,\text{lb}}\) and \(t_{f,\text{ub}}\) may provide better solutions. Figure \ref{fig: traj eci variable} shows the trajectory for the best solution in the J2000 ECI frame. Figure \ref{fig: ecl duration variable} shows the duration of each of the eclipses for the solution. An interesting aspect of the solution is that Moon eclipses occur in this transfer solution, with one Moon eclipse  occurring very close to the NRHO insertion being the limiting eclipse duration.

It is hypothesized that it took longer to reach convergence because of the sensitivity introduced by making \(t_f\) a decision variable. Small changes in \(t_f\) can potentially cause large changes in the initial conditions, \(\bm{x}(t_f)\), depending on how close \(\bm{x}(t_f)\) is to perilune. This means that changes to \(t_f\) itself would require changes to the other parameters for the solution to converge. However, in the current formulation, all parameters are being optimized at the same level. It is hard to fully characterize this hypothesis without performing many runs of PSO, since PSO is a stochastic optimization algorithm. One future work is to investigate optimizing \(t_f\) using a bi-level optimization algorithm. Nonetheless, considering \(t_f\) as a design variable shows that it is possible to obtain solutions with a better time of flight within a new departure window. A potential use of our proposed formulation is to make the bounds \(t_{f,\text{lb}}\) and \(t_{f,\text{ub}}\) large to find a departure window over a wide range of time, or the bounds can be made small to improve optimality once an initial solution is obtained.

\section{Conclusion}
We presented a methodology for efficiently finding low-thrust spacecraft transfer trajectories  under constant acceleration from a geostationary transfer orbit (GTO) to the near-rectilinear halo orbit (NRHO) earmarked for NASA's Gateway. The method is based on a closed-loop control law derived from a novel parameterization of quadratic control-Lyapunov functions. This control law is applied in a backward-in-time sense to generate solutions departing from the GTO and inserting into the NRHO. Solutions may also be obtained that depart from the NRHO and insert into the GTO. 

To solve the resulting trajectory optimization problems, the parameters of the control law, time of flight, and NRHO insertion date are all optimized simultaneously with a stochastic optimization algorithm -- particle swarm optimization (PSO). The cost function is designed to prioritize 1) convergence to the target orbit, 2) satisfaction of the constraint that all eclipse durations be less than 90 minutes, and 3) minimization of the time of flight. Results indicate that eclipse-feasible solutions can be obtained on the order of 10 minutes with the processing power of a personal laptop computer. Solutions obtained can serve as high-quality initial guesses to NASA's high-fidelity trajectory optimization tools such as Copernicus and GMAT.



\section{Acknowledgment}
We would like to acknowledge Saeid Tafazzol for his useful suggestions and discussions on efficiently parameterizing \(n\)-dimensional positive-definite matrices.




\bibliographystyle{AAS_publication}   
\bibliography{LowThrust}   

\end{document}